\documentclass[12pt]{article}
\pdfoutput=1
\usepackage[a4paper]{geometry}
\usepackage{jheppub, amsmath,amssymb,amsfonts,amsxtra, mathrsfs, makeidx,graphics,graphicx,amsthm,epsfig, bm,longtable,float, color,tikz,mathtools,xfrac,footnote,rotating, lscape, makecell, environ,mathtools, empheq}

\usepackage{amsmath,amscd,amstext,amsbsy,amsopn,amsthm,amsxtra,upref,amsfonts,amssymb,euscript,latexsym,graphicx,color,braket,slashed,multicol,mdframed,tikz,tikz-3dplot,todonotes,hyperref,pgfplots,mathrsfs,bbm,array,cleveref,graphicx,wrapfig,lscape,youngtab,bbold,float,multirow,longtable,rotating,epstopdf,tikz-cd,parskip}  
 
\usepackage[T1]{fontenc} 

\usepackage{mathtools}
\usepackage{colortbl}
\usepackage{nicematrix}
\usepackage{diagbox}

\usepackage{subfig}
\usepackage{multirow}
\usepackage{adjustbox}

\usepackage{amsthm}
\usepackage{hyperref}

\usepackage{amsmath}
\usepackage{amsfonts}
\usepackage{commath}
\usepackage[english]{babel}
\usepackage{setspace}
\usepackage{youngtab}

\usepackage{epsfig,amssymb} 
\usepackage{amsmath}

\usepackage{amscd,color}
\usepackage{amsmath,graphicx}
\usepackage{verbatim,booktabs}

\usepackage{latexsym}
\usepackage{amsmath,amsfonts,amssymb,amsthm}
\usepackage{amsmath,amsthm}

\newtheorem{claim}{Claim}
\renewcommand{\theclaim}{\Alph{claim}}

\DeclareMathOperator{\Pf}{Pf}

\newcommand{\zz}{\mathbb{Z}}
\newcommand{\rr}{\mathbb{R}}
\newcommand{\cc}{\mathbb{C}}
\newcommand{\nn}{n}
\newcommand{\NN}{\mathcal{N}}

\newcommand{\Gammal}{\Gamma^{\text{line}}}
\newcommand{\Gammam}{\Gamma^{\text{max}}}

\newlength\mylen

\usepackage{tikz}
\usetikzlibrary{tikzmark}
\usetikzlibrary{math} 
\usetikzlibrary{matrix}
\usetikzlibrary{calc}
\usetikzlibrary{positioning}
\definecolor{azzurro}{RGB}{0,118,186}
\definecolor{rosso}{RGB}{255,0,0}

\usepackage{array}

\graphicspath{ {./images/} }

\definecolor{mandarancio}{RGB}{255,147,0}
\definecolor{azzurro}{RGB}{0,118,186}
\definecolor{terrabruciatadiconcorezzo}{RGB}{181,23,0}

\def\centerarc[#1](#2)(#3:#4:#5)
{ \draw[#1] ($(#2)+({#5*cos(#3)},{#5*sin(#3)})$) arc (#3:#4:#5) }

\def\matSL[#1,#2;#3,#4] 
{	\left( \begin{array}{cc} #1 & #2 \\ #3 & #4 \end{array} \right)}

\def\pf[#1] {\text{Pf} \left( #1 \right)}

\newcommand{\munutable}[6] 
{
\begin{equation}
\begin{array}{|c|l|l|}
\hline r=#2 & \multicolumn{2}{c|}{#1}\\
\hline \nu \backslash \mu & \multicolumn{1}{c|}{1} & \multicolumn{1}{c|}{-1} \\
\hline 1 &\begin{array}{l} #3 \end{array} &\begin{array}{l} #4 \end{array} \\
\hline-1 & \begin{array}{l} #5 \end{array} & \begin{array}{l} #6 \end{array} \\
\hline
\end{array}
\end{equation}
}

\def\strata(#1){ \mathcal{S}^{(#1)} }
\def\DD(#1,#2){\left\langle #1, #2 \right\rangle}

\def\SFoldNotation{2}
\if\SFoldNotation1
	\def\sfold(#1,#2){ #1 S_{#2}\text{-fold} }
	\def\sfolds(#1,#2){ #1 S_{#2}\text{-folds} }
\fi
\if\SFoldNotation2
	\def\sfold(#1,#2){ S_{#2}\text{-fold} \text{ of type }#1}
	\def\sfolds(#1,#2){ S_{#2}\text{-folds}  \text{ of type }#1}
\fi

\def\algstate[#1,#2] { \left| #1 ; \, #2 \right\rangle}
\def\longstate[#1,#2]{\overline{ \left| #1 , #2 \right\rangle}}

\usepackage{booktabs}
\title{
\begin{center}
Exceptional S-fold SCFTs are almost trivial
\end{center}
}

\author[a]{Antonio 
Amariti}
\author[a,b]{and Simone 
 Rota}
\affiliation[a]{INFN, Sezione di Milano, Via Celoria 16, I-20133 Milano, Italy}

\affiliation[b]{Dipartimento di Fisica, Universit\`a degli Studi di Milano, Via Celoria 16, I-20133 Milano, Italy}
\emailAdd{antonio.amariti@mi.infn.it,  simone.rota@mi.infn.it}

\abstract{
We study 
4d exceptional S-fold SCFTs obtained from the 6d $(2,0)$ theories of type $E_{6,7,8}$. We show that all but one of these theories are discrete gaugings of free theories because they do not admit a consistent charge lattice.
We compute the 1-form symmetry of the 
 only interacting theory, the $k=4$ exceptional S-fold SCFT of type $E_8$, and find that it is trivial.
Along the way we develop a consistency condition for the Coulomb Branch  stratification of $\mathcal{N}=2$ SCFTs with characteristic dimension $\varkappa \neq \{1,2\}$ and show that the triviality of (most) exceptional S-fold SCFTs follows directly from this constraint.
}

\begin{document}

\maketitle

\pagebreak

\section{Introduction}
\label{sec:introduction}

The role of conformal field theories (CFTs) as fixed points of the RG flow is a basic building block of our understanding of quantum field theories.
A large and interesting class of interacting CFTs can be obtained from string/M-theory, either through dimensional compactification or geometric engineering. Many of these theories lack a conventional Lagrangian description, therefore the study of their dynamics should involve an analysis of their stringy construction, aided by field theoretical constraints, for example those originating by symmetries. 

When, on top of the conformal symmetry, a CFT also enjoys supersymmetry then field theoretical results can strongly constrain a theory. As an example it is widely believed that in 4d with 16+16 conserved supercharges all the CFTs are classified by $\NN=4$ SYM theories with arbitrary gauge group, possibly with the addition of topological terms in the action. With a lower amount of supersymmetry such a complete classification is not available, although promising progress has been made in the last two decades for SCFTs with $\NN\geq2$ SUSY \cite{Argyres:2015gha,Argyres:2016xmc,Argyres:2016xua,Argyres:2015ffa,Martone:2021ixp,Kaidi:2021tgr,Xie:2015rpa,Chen:2016bzh,Wang:2016yha,Chen:2017wkw,Xie:2021hxd,Cecotti:2021ouq,Martone:2020nsy,Argyres:2020wmq,Martone:2021drm,Argyres:2020nrr,Caorsi:2018zsq,Argyres:2018zay,Caorsi:2018ahl,Caorsi:2019vex,Argyres:2019yyb,Argyres:2018wxu,Bourget:2018ond,Argyres:2019ngz,Argyres:2017tmj,Nishinaka:2016hbw,Argyres:2023eij}. 
An important ingredient that renders a classification program feasible is the existence of a Coulomb Branch  (CB), an $r$-complex-dimensional space of vacua, with $r$ the rank of the SCFT, where on general points the low energy dynamics is that of a $\NN\geq2$ $U(1)^r$ gauge theory where all the charged states are massive.
On non-general singular points of the CB some charged states become massless and give rise to non-trivial dynamics in the IR. 
Then the analysis of the interesting physics of $\NN\geq2$ SCFTs boils down to what happens at singularities of the CB.
The theory arising on a codimension-$n$ singularity is a theory with rank $n<r$, making it possible to study $\NN\geq2$ SCFTs “by induction” on the rank: the properties of a rank-$r$ theory are related to the properties of the theories supported on its singularities, which have rank less than $r$. 
This procedure has been referred as CB stratification \cite{Argyres:2019yyb} and in the following we will borrow this terminology.
In this paper we apply this general idea to the charge lattice $\Gamma$ of $\NN\geq2$ theories which is the lattice of electromagnetic charges under $U(1)^r$ of the massive states in a generic vacua of the CB, with the associated Dirac pairing $J$.

The study of the charge lattices intertwines with the study of the generalized symmetries of the SCFT \cite{Gaiotto:2014kfa}, as seen for example from the analysis of 1-form symmetries in lagrangian theories \cite{Aharony:2013hda}. 
In $\NN=2$ theories the charge lattice is well defined even in the absence of a lagrangian description and this relationship has been exploited for example in \cite{DelZotto:2022ras, Argyres:2022kon, Amariti:2023hev, Closset:2023pmc}.
In particular in $\NN=2$ theories there is a close connection between the charge lattice $\Gamma$ and the 1-form symmetry group $G^{(1)}$. 
Indeed the objects charged under 1-form symmetries, which are Wilson-'t Hooft lines \cite{Kapustin:2005py} in a generic CB vacua, are constrained by the spectrum of charged local states through the Dirac quantization condition.
More precisely, given a basis of the charge lattice, the Dirac pairing matrix $J$ in this basis is related to the order of $G^{(1)}$ as follows \cite{Argyres:2022kon}:
\begin{equation}
\left| G^{(1)} \right| = \left| \pf[J] \right| \, .
\end{equation}
 2-form symmetries, on the other hand, are related to discrete gauging \cite{Gaiotto:2014kfa}.
In 4d gauging a discrete 0-form symmetry generates a magnetic 2-form symmetry, whose topological operators are the Wilson lines of the discrete gauge group. On top of that, if the 0-form symmetry acts non-trivially on the CB, this gauging generates singularities at the CB points that are fixed under the action of the symmetry. There are no states becoming massless at these new singularities, therefore there are no BPS states with vanishing central charge  there. It then follows that 2-form symmetries can be used as indicators of discrete gauging, and the same idea applies also for CB singularities with no massless charged state.
Summarizing, in $\NN\geq2$ SCFTs the structure of higher form symmetries gives information about the local dynamics, namely on the charge lattice and BPS condition of charged states, and viceversa. 

Motivated by this discussion we study a class of 4d SCFTs with $\NN=3$ SUSY denoted as exceptional S-fold theories, first constructed in \cite{Garcia-Etxebarria:2016erx} (see also \cite{Aharony:2015oyb} for general properties of $\NN=3$ SCFTs). 
We analyze the structure of higher-form symmetries from the analysis of the charge lattice and possible  discrete gaugings for each case. 
It is challenging to study CFT data from the stringy definition of such theories, a notable exception is the CB geometry computed in \cite{Kaidi:2022lyo}. Our analysis heavily relies on the approach and results of \cite{Kaidi:2022lyo}.

More technically $\NN=3$ S-fold SCFTs are
 labelled by an integer $k=3,4,6$, called the order of the S-fold, and a simply-laced Lie algebra $\mathfrak{g}$, and we will denote them as “$S_k$-folds of type $\mathfrak{g}$”.
 The 
 $\sfolds(A_r,k)$, denoted here as “regular” S-folds, engineer the theories of \cite{Garcia-Etxebarria:2015wns}, and are equivalent to Type IIB setups.
The 1-form symmetries of regular S-fold theories were computed in \cite{Etheredge:2023ler, Amariti:2023hev}
and the 2-form symmetries and possible discrete gauging were analyzed in \cite{Aharony:2016kai}.
We will use regular S-fold SCFTs as a testing ground for developing our prescriptions.
The 
$\sfolds(D_r,k)$ and $\sfolds(E_r,k)$ are called “exceptional” S-fold theories. In this paper we only discuss the $\mathfrak{g}=E_{6,7,8}$ case, but our procedures can be straightforwardly applied to the 
$\sfolds(D_r,k)$ as well. 

We thus consider a total of nine theories, the $\sfolds(E_{6,7,8},k)$, which are so far candidates for being interacting $\NN=3$ SCFTs with rank varying from 2 to 4. 
We show that all but one of these theories, the $\sfold(E_8,4)$, are discrete gauging of free theories. This is essentially due to the fact that they do not admit a  charge lattice consistent with the CB stratification, the Dirac quantization condition as well as the constraints on the central charges by  \cite{Martone:2020nsy}.
In this paper we will denote a charge lattice as consistent if it satisfies each of these three conditions, while we will denote as inconsistent the lattices that fail to satisfy one or more of them.
Exceptional S-folds can be considered sporadic even when compared to the “regular” $\sfolds(A_r,k)$, nevertheless we find that the obstruction to having a consistent charge lattice that they exhibit generalizes nicely to all $\NN=2$ SCFTs with characteristic dimension\footnote{The characteristic dimension $\varkappa$, introduced in \cite{Cecotti:2021ouq}, is an invariant of $\NN=2$ SCFTs that can be computed from the scaling dimensions of the CB invariants.}
 $\varkappa$ different from 1 or 2 \cite{Cecotti:2021ouq}.

The rest of this paper is organized as follows: in Section \ref{sec:Sfolds} we outline our procedure by generalizing the known case of $\sfolds(A_r,k)$. We review the computation of the CB of S-folds by  \cite{Kaidi:2022lyo} and we give formulae for computing the charge lattice. We analyze the relation between possible discrete gaugings and the presence of BPS charged states and we apply it to show that some strings that cross “regular” S-folds do not produce BPS states. In Section \ref{sec:exceptional} we study the main theories of interest of this paper, the exceptional $\sfolds(E_{6,7,8},k)$. 
We compute their charge lattices and show that most of them are inconsistent.
In Section \ref{sec:bound} we apply the ideas developed in the rest of the paper to the general case of $\NN=2$ SCFTs with characteristic dimension $\varkappa \neq \{1,2\}$ \cite{Cecotti:2021ouq}. We obtain a consistency condition on the CB stratification of these theories and an upper bound for the order of the 1-form symmetry group in the rank-2 case.
For the sake of readability, before delving into the main body of this paper, we find it useful to sketch our procedure and present our results.

\subsection{General strategy}
Our approach to the study of charge lattices of S-fold theories boils down to two main ideas. The first idea is based on the results of \cite{Kaidi:2022lyo}, where it was shown that the moduli space of an S-fold SCFT of type $\mathfrak{g}$ can be obtained as a slice of the moduli space of a “parent” $\NN=4$ SYM theory with gauge algebra $\mathfrak{g}$. In analogy with this result, we compute the charge lattice of an S-fold theory as a sublattice of the charge lattice of the parent $\NN=4$ SYM. 

The second ingredient is the consistency of the structures of $\NN=2$ SCFTs along the CB stratification. In our case this boils down to the fact that the charges that become massless at some codimension-$n$ singularity on the CB must generate the charge lattice of some rank-$n$ theory supported on the singularity. If this is not the case, then the singularity can not support an interacting theory and must be empty. The singularity itself then supports a discrete gauging of a free theory, and the SCFT can be considered as a discrete gauging of a parent theory. 
In most exceptional $\sfolds(E_{6,7,8},k)$  we find that none of the codimension-1 singularities can support an interacting theory, signaling that the exceptional S-fold theory itself is a discrete gauging of a free $\NN=4$ theory. 

This procedure is particularly powerful when considering the codimension-1 singularities of a maximally strongly coupled theory. If the singularity is non-empty then it must support a rank-1 $\NN=2$ SCFT. We have a full classification of these theories \cite{Argyres:2016xmc, Argyres:2015ffa, Argyres:2015gha, Argyres:2016xua}
and their charge lattices are characterized by the absolute value of the Pfaffian of the Dirac pairing $J$ \cite{Argyres:2022kon}:
\begin{equation}	\label{eq:introPf}
\left| \pf[J] \right| = \begin{cases}2 & \text { (discrete gauging of) } \mathcal{N}=2^* SU(2) \mathrm{SYM} \\ 1 & \text { otherwise }\end{cases}
\end{equation}
For any other values of $\left| \pf[J] \right|$ on a codimension-1 singularity, the corresponding states can not be BPS and the singularity must be empty. 

Given an exceptional S-fold theory our analysis roughly follows these steps:
\begin{itemize}
\item Determine the CB  geometry as in \cite{Kaidi:2022lyo}.
\item Compute the charge lattice and Dirac pairing from the parent theory.
\item Compute the sublattice of charges that should become massless on all codimension-1 singularities.
\item If these lattices are compatible with one of the options in \eqref{eq:introPf} then there is a SCFT supported there, otherwise the singularity is empty.
\item Impose the constraints on the central charges from  \cite{Martone:2020nsy}.
\end{itemize}
At the end of these steps if there are some singularities which support an interacting SCFT we claim that the S-fold SCFT is non-trivial. Instead, if all the singularities are empty, we claim that the S-fold theory is a discrete gauging of a free theory.

\subsection{Results}
\begin{table}
\centering
\begin{tabular}{|c|c|c|c|}
\hline
\diagbox{$k$}{$\mathfrak{g}$}
 & $E_6$ & $E_7$ & $E_8$ \\
 \hline
 3 & 
 \cellcolor{red!50}
 $\begin{array}{c} 
 \mathcal{C} = \cc^3/G_{25} \\ 
 \text{d.g. of $U(1)^3$ $\NN=4$} \\ 
 \end{array}$
 & 
    \cellcolor{orange!50}
 $\begin{array}{c} 
 \mathcal{C} = \cc^3/G_{26} \\ 
 \text{d.g. of $U(1)^3$ $\NN=4$} \\ 
 \end{array}$
 & 
  \cellcolor{red!50}
 $\begin{array}{c} 
 \mathcal{C} = \cc^4/G_{32} \\ 
 \text{d.g. of $U(1)^4$ $\NN=4$} \\ 
 \end{array}$\\ 
 \hline
  4 &
     \cellcolor{orange!50}
   $\begin{array}{c} 
 \mathcal{C} = \cc^2/G_{8} \\ 
 \text{d.g. of $U(1)^2$ $\NN=4$} \\ 
 \end{array}$
  &
     \cellcolor{orange!50}
   $\begin{array}{c} 
 \mathcal{C} = \cc^2/G_{8} \\ 
 \text{d.g. of $U(1)^2$ $\NN=4$} \\ 
 \end{array}$
  & 
  $\begin{array}{c} 
 \mathcal{C} = \cc^4/G_{31} \\ 
 \text{Interacting SCFT} \\ 
 \text{$G^{(1)}=\mathbb{1}$, $12c=372$ } 
 \end{array}$\\ 
   \hline
  6 & 
   \cellcolor{red!50}
  $\begin{array}{c} 
 \mathcal{C} = \cc^2/G_{5} \\ 
 \text{d.g. of $U(1)^2$ $\NN=4$} \\ 
 \end{array}$ 
  &
   \cellcolor{orange!50}
   $\begin{array}{c} 
 \mathcal{C} = \cc^3/G_{26} \\ 
 \text{d.g. of $U(1)^3$ $\NN=4$} \\ 
 \end{array}$
  & 
   \cellcolor{red!50}
  $\begin{array}{c} 
 \mathcal{C} = \cc^4/G_{32} \\ 
 \text{d.g. of $U(1)^4$ $\NN=4$} \\ 
 \end{array}$\\
  \hline
\end{tabular}
\caption{Properties of exceptional S-folds of type $\mathfrak{g}=E_n$. For each theory the CB  $\mathcal{C}$ is reproduced from \cite{Kaidi:2022lyo} and we specify wether the theory is a discrete gauging (d.g) of a free theory. Red cells are theories whose CB  
\colorbox{red!50}{do not admit any consistent charge lattice}, 
orange cells are theories where the
\colorbox{orange!50}{charge lattice is incompatible with the constraints on the central charges}.
 For the only interacting SCFT the 1-form symmetry group $G^{(1)}$ is written and the central charge is reproduce from \cite{Kaidi:2022lyo}.}
 \label{tab:results}
\end{table}

We find that all but one of the exceptional S-fold SCFTs of type $E_r$ are discrete gauging of free theories, the exception being the S-fold of type $E_8$ with $k=4$, also called the $G_{31}$ theory. In particular, the S-fold theories of type $E_6$ and $E_8$ with $k=3,6$ do not have consistent charge lattices. In these theories, on any codimension-1 singularity the charges that should become massless span a rank-2 sublattice where the Dirac pairing is such that $\left| \pf[J^{(1)}] \right| = 3$, and comparing with eq. \eqref{eq:introPf} there is no rank-1 SCFT that can be supported there. Therefore all codimension-1 singularities are empty, and the S-fold theories themselves are discrete gauging of free theories.
All S-fold theories of type $E_7$ and the S-fold theory of type $E_6$ with $k=4$ admit a consistent charge lattice, but this lattice is incompatible with the constraints coming from the central charge formulae of \cite{Martone:2020nsy}. 

The only S-fold theory of type $E_{6,7,8}$ that has a well defined charge lattice compatible with the formulae of \cite{Martone:2020nsy} is the $G_{31}$ theory. We claim that this is an interacting SCFT. The theory has rank equal to 4 and the CB  and central charges are those computed in \cite{Kaidi:2022lyo}, see Table \ref{tab:results}. Furthermore we find that the 1-form symmetry group of this theory is trivial.

By applying similar ideas to $\NN=2$ SCFTs with characteristic dimension $\varkappa \neq\{1,2\}$ we find the following:
\renewcommand{\theclaim}{\Alph{claim}}
\begin{claim}
The order of the 1-form symmetry group $G^{(1)}$ of an $\NN=2$ rank-2 SCFT with $\varkappa \neq \{1,2\}$ satisfies $1\leq \left|G^{(1)}\right| \leq 4$. The upper bound can only be saturated by stacks of lower rank theories.	\label{ClaimA}
\end{claim}

\begin{claim}
An $\NN=2$ SCFT with $\varkappa\neq\{1,2\}$ and rank $r\geq2$  that is not a stack of lower rank theories must have at least one codimension-1 singularity that supports (a discrete gauging of) $\NN=2^*$ $SU(2)$ SYM.	\label{ClaimB}
\end{claim}

We show that our results regarding  $\sfolds(E_{6,7,8},k)$, namely that most of them are discrete gauging of free theories, boils down to the constraint of \textbf{Claim} \ref{ClaimB}, possibly in conjunction with the constraints on the central charges given by the formulae of \cite{Martone:2020nsy}.

\section{S-folds SCFTs}
\label{sec:Sfolds}
In this Section we outline our procedure for analyzing various properties of $\NN=3$ S-fold theories. We do so by studying explicit examples of S-folds SCFTs engineered in Type IIB \cite{Garcia-Etxebarria:2015wns}, which we denote as “regular” S-folds, providing various prescriptions that will apply to the general cases of exceptional S-folds \cite{Garcia-Etxebarria:2016erx} engineered in M-theory discussed in Section \ref{sec:exceptional}. 
All the results contained in this Section have already appeared in the literature and most of the techniques are well known, with the exception of the discussion given in Subsection \ref{sebsec:2form}. There we leverage the stratification of the CB  and the classification of $\NN=2$ rank 1 SCFTs to constrain the BPS spectrum and ultimately understand the 2-form symmetries of these theories. This argument, to the best of our knowledge, is original.

\begin{table}[h!]
	\centering
	\begin{tabular}{c|cccc}
		\toprule
		$SL(2,\zz)$ & $S^2=-\mathbb{I}_2$ & $(ST)^{-1}$ & $S$ &$(S^3 T)^{-1}$ \\ \midrule
		$k$ & $2$ & $3$ & $4$ & $6$ \\ 
		$\rho_k$ & $\begin{pmatrix}
			-1 & 0 \\ 0 & -1
		\end{pmatrix}$ & $\begin{pmatrix}
			0 & 1 \\ -1 & -1
		\end{pmatrix}$ & $\begin{pmatrix}
			0 & -1 \\ 1 & 0
		\end{pmatrix}$ & $\begin{pmatrix}
			0 & -1 \\ 1 & 1
		\end{pmatrix}$  \\
		$\tau$ & any $\tau$ & $e^{2i\pi/3}$ & $i$ & $e^{2i\pi/3}$ \\ 
		\bottomrule
	\end{tabular} \caption{Elements $\rho_k$ of $SL(2,\zz)$ of order $k$ used in $S$-fold projections, and the corresponding fixed coupling $\tau$.}
	\label{tab:rho}
\end{table}

We begin with a quick review of S-folds in Type IIB.
S-fold SCFTs can be engineered in Type IIB string theory as the low energy theory on the worldvolume of a stack of $\nn$ D3-branes that probe an S-fold singularity (see \cite{Garcia-Etxebarria:2015wns} for details). This singularity is obtained by a $\zz_k$ quotient of Type IIB which involves both a spacetime orbifold and an S-duality action, which becomes a symmetry for particular values of the axiodilaton. The spacetime orbifold is $\rr^{3,1} \times (\cc^3/\zz_k)$, where D3-branes are extended along $\rr^{3,1}$, and the S-duality action is given by an element $\rho_k \in SL(2,\zz)$ of the S-duality group of Type IIB. One can think about this non-geometric spacetime as follows: looping around a cycle in  $\cc^3/\zz_k$ every object in string theory is acted upon by the S-duality transformation $\rho_k$. 
This Type IIB non-geometric singularity can alternatively be described by a geometric singularity in F-theory, where the F-theory torus has a  $\rho_k$ monodromy around the $\cc^3/\zz_k$ singularity. The F-theory picture will not be relevant in this paper, and we refer the reader to the original literature on this topic \cite{Garcia-Etxebarria:2015wns,Aharony:2016kai}. 

The S-duality element $\rho_k$ must generate a $\zz_k$ subgroup of $SL(2,\zz)$, which is only possible for $k=1,2,3,4,6$. 
Furthermore, the axiodilaton $\tau$ must be fixed by $\rho_k$ in order for the subgroup generated by $\rho_k$ to be a symmetry of the theory.
The S-duality elements  $\rho_k$ with the corresponding values of $\tau$ are listed in Table \ref{tab:rho}.
In the absence of the S-fold the stack of D3 brane preserves sixteen supercharges in 4d: $Q_i$, $i=1,2,3,4$\footnote{Each $Q$ is a four dimensional Dirac spinor with four components. }.
The S-duality transformation $\rho_k$ acts on the supercharges as \cite{Kapustin:2006pk}:
\begin{equation}
	\rho_k: Q_i \to e^{\frac{\pi i}{k}} Q_i \qquad i=1,2,3,4
\end{equation}
On the other hand the spacetime orbifold corresponds to an R-symmetry transformation $r_k \in SU(4)_R$ and can be chosen such that its action  on the supercharges is:
\begin{equation}
r_k: \left\{
\begin{split}
	&Q_i \to e^{-\frac{\pi i}{k}} Q_i \qquad i=1,2,3
	\\
	&Q_4 \to e^{\frac{3 \pi i}{k}} Q_4
\end{split}
\right.
\end{equation}
Under the combined action $\rho_k \cdot r_k$ the supercharges $Q_{1,2,3}$ are preserved while $Q_4$ transforms as:
\begin{equation}
	\rho_k \cdot r_k: \, Q_4 \to  e^{\frac{2 \pi i}{k}} Q_4
\end{equation}
For $k=1,2$ this supercharge is preserved as well and the resulting 4d theory has $\NN=4$ supersymmetry. The case $k=1$ corresponds to no projection at all, and engineers $\mathfrak{su}(N)$ $\NN=4$ SYM, while the case $k=2$ corresponds to the orientifold plane O3 and engineers $\NN=4$ SYM with gauge algebra $\mathfrak{d}_n, \mathfrak{b}_n$ or $\mathfrak{c}_n$ depending on the discrete torsion to be discussed briefly.
The cases of interest in this paper are $k=3,4,6$ where generally only twelve supercharges are preserved and the low energy theory on the stack of D3-branes is an $\NN=3$ SCFT. 

It was shown in \cite{Aharony:2016kai} that generally one has the possibility to introduce a discrete torsion in the S-fold background, that is a non-trivial flux for the Type IIB 2-form fields $B_2$ and $C_2$ around a non-contractible 2-cycle of the holographic background $AdS_5 \times (S^5/\zz_k)$. The 2-form fields transform in the two dimensional representation of the S-duality group $SL(2,\zz)$, therefore their flux on this 2-cycle is classified by the second twisted cohomolgy groups $H_2(AdS_5 \times (S^5/\zz_k); (\zz \oplus \zz)_{\rho_k}) = H_2(S^5/\zz_k; (\zz \oplus \zz)_{\rho_k})$. These groups were computed in \cite{Aharony:2016kai}, see also \cite{Etheredge:2023ler} for a review. One finds:
\begin{equation}
H_2\left(S^5/\zz_k; (\zz \oplus \zz)_{\rho_k} \right) =
\left\{
\begin{split}
	&\zz_2 \times \zz_2 ,
	\\
	&\zz_3,
	\\
	&\zz_2,
	\\
	&\mathbb{1},
\end{split}
\qquad\quad
\begin{split} k=2 \\ k=3 \\ k=4 \\ k=6 \end{split}
\right.
\end{equation}
where $\mathbb{1}$ is the trivial group. Therefore there are four choices of discrete torsion for the orientifold ($k=2$) corresponding to the O3$^-$, O3$^+$, $\widetilde{\text{O3}^-}$ and $\widetilde{\text{O3}^+}$ orientifold planes respectively. For $k=3$ there are three choices, one with trivial discrete torsion and two with non-trivial discrete torsion.
The two choices with non-zero discrete torsion are related by charge conjugation so there are only two physically different choices: trivial or non-trivial discrete torsion. Finally for $k=4$ one can have trivial or non-trivial discrete torsion and for $k=6$ the only choice is to have trivial discrete torsion.

In summary the S-fold setup of \cite{Garcia-Etxebarria:2015wns}, briefly reviewed above, gives rise to an infinite family of $\NN=3$ SCFTs parametrized by the number $r$ of D3-branes, the order of the quotient $k$ and, when allowed, the choice of discrete torsion.
There are five variants of $\NN=3$ S-folds which we denote as $S_{k,\ell}$ following the notation of \cite{Aharony:2016kai}. Here $\ell=1$ corresponds to the absence of discrete torsion and $\ell=k$ corresponds to non-trivial discrete torsion. The five variants are therefore $S_{3,1}, S_{4,1}, S_{6,1}, S_{3,3}$ and $S_{4,4}$.

This concludes our brief review of S-folds in Type IIB, in the remainder of this section we will review some properties of the corresponding SCFTs, namely the moduli space, charge lattice, 1-form and 2-form symmetries.
For a more in-depth analysis of the string theory setup we refer the reader to the original literature \cite{Garcia-Etxebarria:2015wns, Aharony:2016kai}.

\subsection{Moduli space}
\label{sec:moduli_space}
The S-fold theories have a moduli space of vacua parametrized by the motion of the $N$ D3-branes on the transverse space $\cc^3/\zz_k$ which is given by $(\cc^3)^N / G(k,1,N)$ \cite{Aharony:2016kai}, where $G(k,1,N)$ is a crystallographic complex reflection group (CCRG). 
By choosing an $\NN=2$ subalgebra of the $\NN=3$ superalgebra the R-symmetry group is broken to $(SU(2)\times U(1))_R$ and the moduli space splits into a $N$-dimensional CB , an $2N$-dimensional Higgs branch and a mixed branch with respect to the choice of subalgebra. Of particular interest in this paper is the CB , where the $U(1)$ R-symmetry is broken and the $SU(2)$ R-symmetry is preserved. In the brane picture the CB  can be identified with the space parametrized by the positions $z_i$ of the $N$ D3-branes on a 1-complex-dimensional slice $\cc/\zz_k$ of the transverse space. 
Here $z_i$ is a complex number that parametrize the position of the $i$-th D3-brane on this slice.
The CB  is then $\cc^N / G(k,1,N)$, where $G(k,1,N)$ is generated by the transformations:
\begin{equation} 	\label{eq:ATidentifications}
\begin{split}
	&z_i \to e^{\frac{2\pi i}{k}} z_i
	\\
	&z_i \leftrightarrow z_j
\end{split}
\qquad 
i,j= 1,\dots, N
\end{equation}
The ring of polynomials in the $z_i$ that are invariant under $G(k,1,N)$ is freely generated, meaning that there are no non-trivial relations between the generators. There are $N$ generators whose degrees are:
\begin{equation}
	\Delta = (k, \; 2k,\;  3k, \dots, \; Nk)
\end{equation}

The analysis of moduli space given so far has relied upon the brane picture for regular S-fold. Such a picture will not be available when we consider the generalization to exceptional S-fold, so an alternative approach is desirable. 
A possible approach was presented in \cite{Kaidi:2022lyo}, where the authors studied the moduli spaces of exceptional S-fold theories. Here we briefly review their results, more details can be found in the original paper. The idea is to start with a stack of $Nk$ D3-branes in flat space. The low energy theory is then $\NN=4$ SYM with gauge algebra $\mathfrak{su}(Nk)$ and CB $(\cc)^{Nk}/\mathcal{W}(\mathfrak{su}(Nk))$ parametrized by the scalars $\mathbf{\Phi}$. Here $\mathcal{W}(\mathfrak{g})$ is the Weyl group of the Lie algebra $\mathfrak{g}$. Introducing an $S_{k}$-fold imposes the identification:
\begin{equation}	\label{eq:Sfold_moduli_raw}
	 w \cdot \mathbf{\Phi} =\mathcal{O}_k \mathbf{\Phi} 
\end{equation}
Here $\mathcal{O}_k$ is the action of the S-fold on the scalars, which is given by the R-symmetry transformation:  $\mathcal{O}_k = e^{2\pi i/k}$. $w$ is the Weyl element corresponding to the permutation of branes that maps each D3-brane to its first image under the S-fold and it is equal to the $N$-th power of the Coxeter element $c$:
\begin{equation}
\begin{split}
w =& c^N
\\
c =& s_1 \cdot s_2 \cdot \dots \cdot s_{Nk-1}
\end{split}
\end{equation}
where $s_i$ is the reflection along the $i$-th simple root. Then \eqref{eq:Sfold_moduli_raw} becomes:
\begin{equation}	\label{eq:Sfold_moduli}
	 c^N \cdot \mathbf{\Phi} = e^{2\pi i/k} \mathbf{\Phi} 
\end{equation}
Notice that \eqref{eq:Sfold_moduli} only requires to know the moduli space of the low energy field theory in the absence of the S-fold, which in this case is $\NN=4$ $SU(Nk)$ SYM, and does not rely on a brane picture. This allowed the authors of \cite{Kaidi:2022lyo} to generalize this procedure to exceptional S-fold where the “parent” $\NN=4$ theory has gauge algebra $\mathfrak{e}_n$ or $\mathfrak{d}_n$. 
This generalization requires a choice of an element $w$ of the Weyl group, we will see that this choice is unique under a technical but reasonable assumption.

Mathematically $\mathbf{\Phi}$ is a point in a space acted upon by a real reflection group (the Weyl group), and  \eqref{eq:Sfold_moduli}  identifies the eigenspace of the element $w$ of the Weyl group (in this case $c^N$) with eigenvalue $e^{2\pi i/k}$. 
The action of the Weyl group on this eigenspace is called a reflection subquotient, and has been studied in generality in the mathematical literature see \cite{CCRG} and references therein for a comprehensive review of this topic. 
Here we report some results on reflection subquotient that are relevant for the study of moduli spaces of S-fold theories. Proofs and discussions regarding these mathematical results can be found in \cite{CCRG}, see \textbf{Theorem 11.24}, \textbf{Corollary 11.25}, \textbf{Theorem 11.28} and \textbf{Theorem 11.38} in that reference. 

\begin{itemize}
\item The rank $r$ of an S$_k$-fold $\NN=3$ theory is given by the number of degrees of CB invariants of the “parent” $\NN=4$ SYM that are divisible by $k$.
\item The CB  of an S$_k$-fold $\NN=3$ theory is $\cc^r/\mathcal{C}$ with $\mathcal{C}$ a complex crystallographic reflection group.
\item The degrees of the generators of the S$_k$-fold CB  invariants are the degrees of $\mathcal{C}$, and are given by the degrees of invariants of the “parent” $\NN=4$ theory that are divisible by $k$. 
\item The codimension-1 singularities in the CB  of the S$_k$-fold $\NN=3$ theory  are given by the intersection of the codimension-1 singularities of the “parent” $\NN=4$ theory with the $\NN=3$ CB .
\end{itemize}

In the case of regular S-folds the “parent” theory is $SU(Nk)$ $\NN=4$ SYM, and the degrees of the generators of CB  invariants are:
\begin{equation}
2,3, \dots, Nk
\end{equation}
There are $N$ degrees that are divisible by $k$:
\begin{equation}
	k, 2k, \dots, Nk
\end{equation}
that correspond to the degrees of the complex crystallographic reflection group $G(k,1,N)$. This is consistent with the brane picture analysis, where the CB  was found to be $\cc^N/G(k,1,N)$. 

When a brane picture is not available one needs to specify the element $w$ of the Weyl of the “parent” $\NN=4$ theory that is involved in the S-fold projection \eqref{eq:Sfold_moduli_raw}. Modulo a technical assumption\footnote{Here, following \cite{Kaidi:2022lyo}, we assume that the rank of the $\NN=3$ theory is the highest possible.} such an element is unique up to conjugation and is characterized by having an $r$-dimensional eigenspace with eigenvalue $e^{2\pi i/k}$. Therefore the analysis of the CB  of exceptional S-fold theories boils down to finding an element $w$ of the Weyl group that has an $r$-dimensional eigenspace with eigenvalue $e^{2\pi i/k}$.


\subsection{Charge lattice and 1-form symmetries}
\label{sec:Sfold1form}

The low energy theory in a generic point of the CB  is a $U(1)^r$ gauge theory with no massless charged states. There are massive states with electric and magnetic charges under the various $U(1)$ factors.
The set of EM charges of these massive states form a $2r$-dimensional lattice $\Gamma$ called the charge lattice. 
This lattice is endowed with a Dirac pairing, an antisymmetric bilinear map $\langle \cdot, \cdot \rangle$ taking values in the integers:
\begin{equation}
\langle \cdot, \cdot \rangle: \Gamma \times \Gamma \to \zz
 \end{equation}
 It is convenient to choose a basis $\{\gamma_i\}_{i=1,\dots, 2r}$ of the charge lattice $\Gamma$:
 \begin{equation}
 \Gamma = \text{Span}_{\zz} \left( \{\gamma_i\}_{i=1,\dots, 2r}\right)
 \end{equation}
 In this basis the Dirac pairing is represented by a $2r\times 2r$ antisymmetric matrix $J$. In an SCFT the Dirac pairing must be non-degenerate, which is equivalent to:
 \begin{equation}
 \text{Pf} (J) \neq 0
 \end{equation}


There is a close relationship between the charge lattice and the 1-form symmetry of an $\NN=2$ SCFT \cite{DelZotto:2022ras,Argyres:2022kon}. 
In a generic point of the CB any Wilson-'t Hooft line $\ell$ must have integer Dirac pairing with respect to the massive charged states $\gamma$:
\begin{equation}	\label{eq:mutual_locality}
	\langle \ell, \gamma \rangle \in \zz \qquad \forall \gamma \in \Gamma
\end{equation}
This stems from the fact that due to the Aharonov-Bohm effect, moving the line with charge $\ell$ around the worldline of a particle with charge $\gamma$ produces a phase $e^{2\pi i \langle \ell, \gamma \rangle}$. 
The phase must be a multiple of $2\pi$ in order for the line operator to be well defined, therefore the Dirac pairing between the line and all the local operators must be integer. 
The set of all charges $\ell$ that satisfy the consistency condition \eqref{eq:mutual_locality} is called the line lattice $\Gammal$ and is a refinement of the charge lattice $\Gamma \subset \Gammal$. Two lines in $\Gammal$ do not necessarily have an integer Dirac pairing between themselves, and therefore generally they can not be simultaneously included in the theory. One has to specify a choice of a maximal sublattice of lines $\Gammam \subset \Gammal$ such that:
\begin{equation}
\langle \ell_1, \ell_2 \rangle \in \zz \qquad \forall \ell_1, \ell_2 \in \Gamma_{\text{max}}
\end{equation}
and no additional line charge can be added to $\Gamma_{\text{max}}$ without breaking this consistency condition. 
Generally there are different choices of $\Gammam$ corresponding to 
different global structures, which are theories with the same local dynamics that only differ in the spectrum of extended line operators.
Once a global structure has been chosen the 1-form symmetry group $G^{(1)}$, sometimes referred to as the defect group $\mathbb{D}^{(1)}$,  can be found as:
\begin{equation}
	G^{(1)} = \frac{\Gamma}{\Gammam}
\end{equation}

 An interesting quantity to study in this regard is the absolute value of the Pfaffian of the Dirac pairing $\Pf(J)$ because it is equal to the order of the 1-form symmetry group:
\begin{equation}
	|\Pf(J)| = \left| G^{(1)} \right|
\end{equation}
Indeed one can show that the Dirac pairing $J$ can be put in the following standard form with a change of basis (see Appendix B of \cite{Argyres:2022kon}):
\begin{equation}
J=\left(\begin{array}{rr} 
& D \\
-D &
\end{array}\right) \quad  \quad D=\operatorname{diag}\left\{d_1, \ldots, d_n\right\}, \quad d_i \in \mathbb{N}
\end{equation} 
and the invariant factors $d_i$ can be chosen such that $d_i | d_{i+1}$. The line lattice in this basis is spanned by $\Gammal = \text{Span}_\zz \left( d_1^{-1} e_1, \dots, d_n^{-1} e_n, d_1^{-1} m_1, \dots, d_n^{-1} m_n \right)$, where $\{e_i\}$ and $\{m_i\}$ are respectively the first $n$ and the last $n$ basis vectors of the charge lattice. For any choice of global structure $\Gammam$ it is true that:
\begin{equation}
	\left|	\frac{\Gammal}{\Gammam} \right|
	=
	\left|	\frac{\Gammam}{\Gamma} \right|
	=
	\prod_{i=1}^{n} d_i
	=
	|\Pf(J)|
\end{equation}
When the Pfaffian of the Dirac pairing is a prime number $p$ the 1-form symmetry group must be $\zz_p$, which is the only abelian group of order $p$\footnote{Higher form symmetry group are always abelian\cite{Gaiotto:2014kfa}.}. When the Pfaffian is not a prime number then the 1-form symmetry group can be any of the groups $\prod \zz_{p_j}$ with $\prod p_j = \left|\Pf(J)\right|$, and in general will depend on the choice of global structure $\Gammam$. 

As already discussed the value of $	|\Pf(J)|$ is an invariant of any $\NN=2$ SCFT that equals the order of the 1-form symmetry group. One can intuitively think of this invariant as a measure of “how spread out” the charge lattice is. Indeed the number of  electromagnetic charges that can be added to the charge lattice $\Gamma$ without breaking the Dirac quantization condition is given by  $	|\Pf(J)|-1$. In this sense, the charge lattice $\Gamma$ can not be arbitrarily dense, because $|\Pf(J)|$ is at least 1. In Section \ref{sec:bound} we will see that for maximally strongly coupled SCFTs the charge lattice can not be arbitrarily spread out either. This idea will be discussed in more generality in Section \ref{sec:bound}. 

\subsubsection*{Charge lattices of S-fold theories}
In this Section we review the computation of the charge lattice of regular S-fold SCFTs of \cite{Garcia-Etxebarria:2015wns} and we give a field theoretic prescription to generalize the analysis to exceptional S-folds. Consider a stack of $N$ D3-branes probing an S$_k$-fold without discrete torsion together with the $(k-1)N$ image D3-branes. The local states of the SCFT are associated to finite length $(p,q)$-strings stretched between the D3-branes, plus their images under the S$_k$-fold. Denote as $|(p,q)\rangle_{i,j}$ a state associated to a $(p,q)$-string between the $i$-th and $j$-th D3-brane. The first image of this string is a $(p',q')$-string stretched between the $\pi(i)$-th and $\pi(j)$-th D3-brane. Here $(p',q')$ are related to $(p,q)$ by the S-duality transformation involved in the S-fold:
\begin{equation}
	(p',q') = \rho_k \cdot (p,q)
\end{equation}
And the $\pi(i)$-th D3-brane is the first image of the $i$-th D3-brane. Let us number the D3-branes such that $\pi(i) = i+N$, and $i\sim i+kN$. The following states are invariant under the S-fold action:
\begin{equation}	\label{eq:Sfold_state}
	\overline{|(p,q)\rangle_{i,j} } = \frac{1}{\sqrt{k}} \sum_{t=0}^{k-1} |(\rho_k)^t \cdot (p,q) \rangle_{\pi^t(i), \pi^t (j)}
\end{equation}


The electromagnetic charges of the state $\overline{|(p,q)\rangle_{i,j} }$  can be written as a $2Nk$-dimensional vector\footnote{The factor $1/\sqrt{k}$ is consistent with the charge lattices of $SO(2N)$ $\NN=4$ SYM and the fluxless S-fold SCFTs \cite{Amariti:2023hev} for $k=2$ and $k=3,4,6$ respectively.}:
\begin{equation}	\label{eq:Sfold_Q}
	Q\left[\overline{|(p,q)\rangle_{i,j} } \right] = \frac{1}{\sqrt{k}}(e_1, \dots, e_{Nk}, m_{1},\dots ,m_{Nk} )
\end{equation}
where $e_i$ and $m_i$ are the electric and magnetic charges under the $i$-th D3-brane, respectively.
The Dirac pairing between two states $\phi$ and $\psi $ with charges $e_i, m_i$ and $e'_i, m'_i$ is then:
\begin{equation}	\label{eq:S-fold_pairing}
	\langle \phi, \psi \rangle = \frac{1}{k} \sum_{i=1}^{Nk} (e_i m'_i - e'_i m_i)
\end{equation}
One can show that despite being represented by $2Nk$-dimensional vectors the set of states invariant under the S-fold action \eqref{eq:Sfold_state} only span a $2N$-dimensional lattice, which is the charge lattice of the rank-$N$ S-fold SCFT.

In order to generalize this analysis to the exceptional S-fold case, let us express the various quantities of the S-fold theory \eqref{eq:Sfold_state},\eqref{eq:Sfold_Q} and \eqref{eq:S-fold_pairing} in terms of field theoretical data of the “parent” $\NN=4$ $SU(kN)$ SYM theory, namely the roots $\alpha_{i,j}$ of  $SU(kN)$ and the Cartan matrix $\mathcal{A}_{SU(Nk)}$. This can be done as follows. The string state $|(p,q)\rangle_{i,j}$ correspond to a dyonic state with electric charge $p$ and magnetic charge $q$ with respect to a root $\alpha_{i,j}$ of $SU(kN)$:
\begin{equation}
|(p,q)\rangle_{i,j} \to |\alpha_{i,j}, (p,q)\rangle
\end{equation}
The S-fold acts with a matrix $\rho_k$ on the electric and magnetic charges $(p,q)$ and acts as a permutation on the indices $i,j$. As discussed in the previous Section the permutation corresponds the the action of the Coxeter element $c$ to the $N$-th power on the root $\alpha_{i,j}$. Suppresseing the indices $i,j$ the S-fold action can be written as:
\begin{equation}	\label{eq:Sfoldaction}
\text{S}_k: \algstate[\alpha, (p,q)] \to \algstate[c^N \cdot \alpha, \rho_k\cdot (p,q)]
\end{equation}
The states invariant under the S-fold \eqref{eq:Sfold_state} can be written in the following form in terms of the states of the “parent” $\NN=4$ theory:
\begin{equation} \label{eq:long_states_FT}
\longstate[\alpha, (p,q)] = \frac{1}{\sqrt{k}}\sum_{t=0}^{k-1} \algstate[c^{tN} \cdot \alpha, (\rho_k)^t \cdot (p,q)]
\end{equation}
 The charge lattice of the S-fold theory is spanned by these states for all choices of root $\alpha \in \Delta\left[SU(Nk)\right]$ and  for any $p,q\in\zz$.

The electromagnetic charge of a state $\overline{|\alpha;(p,q)\rangle}$ is given by;
\begin{equation}	\label{eq:Q_E6}
Q\left[\overline{|\alpha;(p,q)\rangle} \right] = 
\frac{1}{\sqrt{k}} \sum_{t=0}^{k-1}
(w \otimes \rho_k )^t \cdot
Q\left[|\alpha; (p,q)\rangle\right]
\end{equation}
where $Q\left[|\alpha; (p,q)\rangle\right]$ is the electromagnetic charge of the corresponding state of  $SU(Nk)$ $\NN=4$ SYM. 
Finally, the Dirac pairing defined on the charge lattice of the S-foldss theory is obtained as a restriction of the Dirac pairing of $SU(Nk)$ $\NN=4$ SYM. Explicitly the Dirac pairing between two states of the S-fold theories with charges $q_i$ and $q_j$ is given by:
\begin{equation}	\label{eq:J_S-fold}
\left\langle q_i, q_j\right\rangle=q_i \cdot J_{SU(Nk)} \cdot q_j^T
\end{equation}
where:
\begin{equation}	\label{eq:J_SUNk}
J_{SU(Nk)}=\left(\begin{array}{cc}
0 & \left(\mathcal{A}_{SU(Nk)}\right)^T \\
-\mathcal{A}_{SU(Nk)} & 0
\end{array}\right)
\end{equation}
is the Dirac pairing of the parent $SU(Nk)$ $\NN=4$ gauge theory (see for example \cite{Argyres:2022kon} for a derivation). 
Then the charge lattice $\Gamma_{N,k}$ of the S$_k$-fold theory is:
\begin{equation}
\Gamma_{N,k}=\left\{Q\left[\overline{|(p, q)\rangle_{i, j}}\right] \mid p, q \in \mathbb{Z}, \quad i, j=1, \ldots, N k\right\}
\end{equation}
and the associated Dirac pairing is given by \eqref{eq:J_S-fold} and \eqref{eq:J_SUNk}.

\subsubsection*{Discrete torsion}
Some S-fold backgrounds can admit a non-zero flux for the Type IIB 2-form fields around cycles of the transverse space. This is possible for $k=2$, giving rise to the orientifold $O3^+$, $\widetilde{O3^+}$ and $\widetilde{O3^-}$, and for $k=3,4$ giving rise to fluxful S-folds denoted as S$_{3,3}$ and S$_{4,4}$. When the discrete torsion is non-zero the S-fold is magnetically charged under the corresponding 2-form field, and strings can end on the S-fold itself. Strings stretched between the S-fold and a D3-brane generate additional states in the SCFT with respect to the fluxless case, and the charge lattice is more dense. 
The states corresponding to strings stretched between the S-fold and a D3-brane can not be written in the form \eqref{eq:long_states_FT}. In order to include them we are led to consider states of the general form:

\begin{equation}	\label{eq:CH_state}
	\overline{\left| \alpha, (p,q) \right>_{\{p_{i,j}\}} } = \frac{1}{\sqrt{k}} \sum_{i=0}^{k-1} \sum_{j=0}^{1}
	p_{i,j} 
\left|c^{ i N} \cdot \alpha ;\left(\rho_k\right)^j \cdot(p, q)\right\rangle
\end{equation}
where $p_{i,j}$ are integers such that the state \eqref{eq:CH_state} are invariant under the S-fold action \eqref{eq:Sfoldaction}.
The sum over $i$ runs from 0 to $k-1$ because $c^{N}$ satisfies its characteristic equation, which is an order $k$ polynomial equation, and similarly the sum over $j$ runs from 0 to 1 because $\rho_k$ satisfies an order 2 polynomial equation.

One can show that such states correctly reproduce the strings stretched between a D3-brane and the S-fold itself. We can therefore see that the presence of a non-trivial discrete torsion can be accounted for by considering the charge lattice spanned by the states \eqref{eq:CH_state} rather than (less general) states \eqref{eq:long_states_FT}. 

In the context of exceptional S-folds the states \eqref{eq:CH_state} will only play a minor role, therefore we will not discuss them further.

\subsection{Discrete gauging and 2-form symmetries}
\label{sebsec:2form}
The S-fold theories obtained from the Type IIB setup can sometimes have a non-trivial 2-form symmetry and can be seen as a discrete gauging of a “parent” theory \cite{Aharony:2016kai}.
Gauging a discrete 0-form symmetry of the parent theory gives rise to a magnetic 2-form symmetry, and viceversa. One can go from the parent theory to the daughter theory by gauging the relative discrete symmetry. This operation is therefore reversible, and one may choose to study either of the two theories without loosing information.
When this is the case it is convenient to study the parent theory itself, for example the Shapere-Tachikawa formula for the central charges is believed to hold only in the absence of 2-form symmetries. 

In this Section we show how to detect 2-form symmetries that arise from the discrete gauging of a 0-form symmetry that acts on the CB . We also give a consistency constraint for the BPS spectrum of $\NN=2$ SCFTs based on the classification of rank-1 theories. We elaborate this analysis in the cases of the O3$^-$ and the flux-less S$_3$-fold. In Section \ref{sec:exceptional} similar considerations will lead us to claim that some exceptional S-fold theories are discrete gaugings of free theories.

\subsubsection*{Strings across the flux-less orientifold}
As a familiar example, consider the O3$^-$ plane, which corresponds to the S-fold with $k=2$ and trivial discrete torsion. The low energy theory on a stack of $N$ D3 branes on top of the O3$^-$ is $\NN=4$ SYM with gauge algebra $\mathfrak{so}(2N)$, and is believed to be a $\zz_2$ discrete gauging of the $\NN=4$ theory with gauge group $Spin(2N)$. Indeed the space parametrized by the transverse motion of the D3-branes is $\mathbb{C}^N/ \left( \mathcal{W}\left[\mathfrak{so}(2N)\right]\rtimes \zz_2\right)$, which is compatible with the moduli space of  $\NN=4$  $Spin(2N)$ with an additional $\zz_2$ identification given by gauging charge conjugation. In this example the “parent” theory has trivial 2-form symmetry and has a $\zz_2$ 0-form symmetry, namely charge conjugation. The theory on the stack of D3-branes is obtained by gauging this $\zz_2$ 0-form symmetry, and therefore has a $\zz_2$ 2-form symmetry.
The 2-form symmetry can be detected by looking at the moduli space $\mathbb{C}^N/ \left( \mathcal{W}\left[\mathfrak{so}(2N)\right]\rtimes \zz_2\right)$. In particular the singularities on moduli space given by the additional $\zz_2$ identifications correspond to configurations where one D3-brane is on top of the orientifold. There are no massless BPS charged states associated to this singularities because the ground state of strings connecting the D3-brane and its image, which have zero length, are projected out by the orientifold. This is consistent with the fact that the $\zz_2$ identification on moduli space is due to a discrete gauging of a $\zz_2$ 0-form symmetry. In general a discrete gauging of a 0-form symmetry that acts non-trivially on the CB  produces singularities where no BPS state becomes massless. Suppose that an S-fold theory $\mathcal{T}$ has a CB :
\begin{equation}
\mathcal{C} = \mathbb{C}^N/\left( \mathcal{G} \rtimes \mathcal{G}' 	\right)
\end{equation}
and charge lattice $\Gamma$. Suppose that on the fixed loci of $\mathcal{G}$ some state $\gamma \in \Gamma$ has zero central charge $\mathcal{Z}$, and therefore becomes massless, while on the fixed points of $\mathcal{G}'$ all the states in the charge lattice are massive. Then $\mathcal{T}$ has a non-trivial 2-form symmetry $G^{(2)} = \mathcal{G}'$ and can be regarded as a $\mathcal{G}'$ discrete gauging of a “parent” theory $\mathcal{T}'$ with CB :
\begin{equation}
\mathcal{C}' = \mathbb{C}^N/\left( \mathcal{G} 	\right)
\end{equation}
and with the same charge lattice $\Gamma$. The “parent” theory $\mathcal{T}'$ has a 0-form symmetry which contains $\mathcal{G}'$ as a discrete subgroup.
Therefore we are able to detect the presence of a non-trivial discrete 2-form symmetry $G^{(2)}$ from the knowledge of the CB  and charge lattice if $G^{(2)}$  arises from a discrete gauging of a 0-form symmetry that acts non-trivially on the CB .

In the example of the O3$^-$ given above the absence of charged massless states on the fixed points of the $\zz_2$ identification can be explained from string theoretical considerations, but one would like a field theoretical argument as well. Consider a point $p$ in $\mathbb{C}^N$ that is fixed under $\zz_2$ and is generic otherwise. 
The prescription given in Section \ref{sec:Sfold1form} to compute the charge lattice $\Gamma$ predicts massless states on this singularity corresponding to $(p,q)$-strings stretched between a D3-brane and its image. Denote as $\Gamma^{(1)}$ the sublattice of $\Gamma$ spanned by these states.
$\Gamma^{(1)}$ should be the charge lattice of a rank-1 QFT $\mathcal{T}^{(1)}$ whose CB  is given by the slice transverse to the singular locus in a neighbor of $p$, namely $\mathbb{C}/\zz_2$.
A basis of $\Gamma^{(1)}$ is given by the states associated to an F1 and a D1 string which we denote as $\psi$ and $\phi$ respectively. The Dirac pairing between these states is $\Pf(J^{(1)} )= \langle \psi, \phi \rangle =4$, therefore they are not mutually local and $\mathcal{T}^{(1)}$ must be an interacting CFT. We have denoted as $J^{(1)}$ the matrix representing the Dirac pairing of the rank-1 theory in this basis.
Furthermore by the argument given in \cite{DelZotto:2022ras,Argyres:2022kon} (see Section \ref{sec:Sfold1form}) $\mathcal{T}^{(1)}$  should have a non-trivial 1-form symmetry group of order 4. 
A full classification of rank-1 $\NN=2$ SCFTs is available\cite{Argyres:2016xmc, Argyres:2015ffa, Argyres:2015gha, Argyres:2016xua}, and a theory such as $\mathcal{T}^{(1)}$ does not exist. In particular the maximum order for the 1-form symmetry group of a rank-1 SCFT is 2 \cite{Argyres:2022kon}, saturated for example by $\NN=4$ $SU(2)$ SYM.
We conclude that the states in $\Gamma^{(1)}$ can not be BPS, therefore on the fixed locus of the $\zz_2$ identification there are no massless states, consistently with the string theory prediction.

\subsubsection*{Strings across the flux-less S$_3$-fold}
We have shown that
the analysis of the charges of states becoming massless on a singularity of the CB  imposes non-trivial constraints on the BPS spectrum of a theory. This is especially interesting to study in non-lagrangian theories, where discrete gaugings and 2-form symmetries are not readily apparent. As an example, we now show that in the flux-less S-folds with $k=3$ the strings stretched between one D3-brane and its image are not BPS. Consequently, these theories are discrete gaugings of other $\NN=3$ “parent” theories, as originally discussed in \cite{Aharony:2016kai}. 
A similar analysis in Section \ref{sec:exceptional} will show that some exceptional S-fold theories, for example the $G_5$ theory discussed in Section \ref{sec:G5}, are discrete gauging of free theories.

The CB  of the regular S$_k$-fold SCFT at rank $r$ theories has two types of codimension 1 singularities on the CB : singularities where two D3-branes coincide and singularities where one D3-brane is on top of the S-fold. When two D3-branes coincide the associated rank-1 theory is always $SU(2)$ $\NN=4$ SYM, which is a consistent rank-1 SCFT, therefore we will focus on the other singularities. When one D3-brane is on top of the S-fold the corresponding rank-1 theory is the rank-1 version of the S-fold theory under consideration. In the case of flux-less S-folds the rank-1 theories are believed to be discrete gaugings of $U(1)$ $\NN=4$, with no massless states charged under the $U(1)$ \cite{Aharony:2016kai}.
 Let us show that this must indeed be the case for the $k=3$ S-fold.
Consider the codimension-1 singularity that arises when the $i$-th D3 brane is on top of the S$_3$-fold.
 In the absence of discrete torsion the charge sublattice  $\Gamma^{(1)}$, associated to the rank-1 theory supported on this singularity, is spanned by $(p,q)$-strings stretched between the $i$-th D3-brane and its image. A possible basis for this lattice is given by the states associated to an F1 and a D1 string, let us denote them as $|f1\rangle$ and $|d1\rangle$ respectively. The Dirac pairing matrix in this basis is:
\begin{equation}
J^{(1)} = \left(
\begin{array}{cc}
0 & \langle f1, d1 \rangle \\ -\langle f1, d1 \rangle & 0
\end{array}
\right)
\end{equation}
The order of the 1-form symmetry group is give by the Pfaffian of the Dirac pairing:
\begin{equation}	\label{eq:DiracSfoldrank1}
|G^{(1)}|=\left| \pf[J^{(1)}]  \right|= \langle f1, d1 \rangle =
\left\{
\begin{array}{c}
3 \quad\quad k=3 \\
2 \quad\quad k=4 \\
1 \quad\quad k=6
\end{array}
\right.
\end{equation}
Where we used \eqref{eq:Sfold_state} and \eqref{eq:Sfold_Q} to compute the charges of $|f1\rangle = \overline{|(1,0)\rangle_{i,i}}$ and $|d1\rangle = \overline{|(0,1)\rangle_{i,i}}$, and we used \eqref{eq:S-fold_pairing} to compute their Dirac pairing.
The CB  of these rank-1 theories is $\mathbb{C}/\zz_k$. For $k=3$ the putative theory on this singularity is inconsistent because, as discussed above, the maximum order for the 1-form symmetry group of a rank-1 $\NN=2$ SCFT is 2. Therefore the states associated to strings stretched between a D3-brane and its images can not be BPS.

The CB  can thus be written as:
\begin{equation}
 	\mathcal{C} = \mathbb{C}^N/\left(G(3,1,r)	\right) =  \mathbb{C}^N/\left(G(3,3,r) \rtimes \zz_{3}	\right) 
\end{equation}
where there are massless charged states on the fixed points of $G(3,3,r)$ and there are no massless charged states on the fixed points of $\zz_3$. This CB  is consistent with the CB  of a $\zz_3$ discrete gauging “parent” theory with CB  $\mathcal{C}'= \mathbb{C}^N/G(3,3,r)$ and 2-form symmetry group $\zz_3$, reproducing the M-theory results of \cite{Aharony:2016kai}. Furthermore we have shown that in a flux-less S-fold background with $k=3$ the states associated to strings stretched between a D3-brane and its image are not BPS, because otherwise the CB  stratification would be inconsistent. This further strengthens the analysis of the BPS spectrum of the rank-2 S-fold theories performed in \cite{Agarwal:2016rvx,Imamura:2016udl}. 
One could perform a similar analysis in the case of flux-full S-folds. The resulting rank-1 theories on the codimension-1 singularities are all consistent in this case, therefore the CB  of these theories is $\mathbb{C}^N/\left(G(k,1,r)	\right)$ and there are massless charged states on all singularities.

\subsubsection*{Strings across the flux-less S$_4$-fold}
As a final example of our techniques before delving into the topic of exceptional S-folds we consider the fluxless S-fold with $k=4$. Similarly to the case of the S$_3$-fold we will show that the strings stretched between a D3-brane and its images do not produce BPS states. In addition to the analysis of the charge lattice we will also consider constraints on the central charges of the theories supported on codimension-1 singularities. These two computations give incompatible results unless the strings across the S$_4$-fold are not BPS and in return the S$_4$-fold SCFTs must be discrete gaugings, reproducing the M-theory results of \cite{Aharony:2016kai}. 
A similar phenomena happens in exceptional S-fold theories, for example the $G_8$ theory discussed in Section \ref{sec:G8} turns out to be a discrete gauging of a free theory.


Consider the rank-$r$ fluxless S$_4$-fold SCFT.
The CB  of the rank-1 SCFT on the singularity that arises when a D3-brane coincides with the S$_4$-fold is parametrized by the motion of the D3-brane on a 1-complex-dimensional slice of the transverse space, and is therefore $\cc/\zz_4$. The order of the 1-form symmetry group of the rank-1 SCFT on this singularity was computed in \eqref{eq:DiracSfoldrank1}:
\begin{equation}
\left|G^{(1)} \right| = \left| \pf[J^{(1)}] \right| =2 
\end{equation}
The only candidate for the rank-1 theory supported on this singularity is the $\NN=3$ preserving $\zz_4$ gauging of $SU(2)$ $\NN=4$ SYM.

We may now consider the central charge of the S$_4$-fold SCFTs. Assuming that the rank-1 theories on all the singularities are not empty and the S$_4$-fold SCFTs are not discrete gaugings the central charges can be computed with the Shapere-Tachikawa formula \cite{Shapere:2008zf}:
\begin{equation}	\label{eq:SfoldST}
2(2 a-c) = \sum_{j=1}^r \Delta_j-\frac{r}{2} = 2 r^2 + \frac{3}{2} r
\end{equation}
where $\{ \Delta_i \} = \{4,8,\dots 4r\}$ are the degrees of invariants of $G(4,1,r)$. The central charges may also be computed using the formulae of \cite{Martone:2020nsy, Martone:2021ixp} that relate data of the rank-1 theories supported on the codimension-1 singularities to the central charge of the rank-$r$ theory:
\begin{equation}	\label{eq:Martone12c} 
12 c=2 r+h_{\mathrm{ECB}}+\sum_{i \in \mathcal{I}} \Delta_i^{\mathrm{sing}} b_i
\end{equation}
where $b_i$ is a quantity associated with the rank-1 theory supported on the $i$-th codimension-1 singularity as follows:
\begin{equation}
b_i:=\frac{12 c_i-h_i-2}{\Delta_i}
\end{equation}
In our theory the extended CB  dimension is  $h_{\mathrm{ECB}} = r$ and the set $\mathcal{I}$ consists of the two codimension-1 singularities. The scaling dimensions $\Delta_i^{\text{sing}}$ of these singularities can be found for example in \cite{Kaidi:2022lyo}, Appendix B. It turns out that the singularity associated to the collision of two D3-branes has scaling dimension $\Delta_1^{\text{sing}}=4 r (r-1)$ and parameter $b_1 = b_{SU(2)} = 3$. The other singularity has scaling dimension $\Delta_2^{\text{sing}}=4r$, therefore \eqref{eq:Martone12c} reduces to:
\begin{equation}
	12c = 3r + 12 r (r-1) + b_2 4r
\end{equation}
Comparing with \eqref{eq:SfoldST} with $a=c$ and solving for $b_2$ one finds:
\begin{equation}
b_2 = \frac{9}{2}
\end{equation}
Which is compatible with having a rank-1 fluxfull S$_4$-fold SCFT on the singularity corresponding to a D3-brane on top of the S$_4$-fold. This is incompatible with the charge lattice computed above, indeed the only possible SCFT on this singularity compatible with the charge lattice is a discrete gauging of $\NN=4$ $SU(2)$ SYM, which would require $b_2=3$. 

We conclude that the states associated to string stretched between a D3-brane and its images do not produce BPS states. Then the rank-1 theory arising when one D3-brane approaches the fluxless S$_4$ fold is not an interacting SCFT, but rather a discrete gauging of free $\NN=4$ Maxwell theory. The S$_4$-fold SCFTs can then be thought as a $\zz_4$ discrete gauging of a parent theory with moduli space $\cc^{3r}/G(4,4,r)$, reproducing the results of \cite{Aharony:2016kai}.

\section{Exceptional S-folds}
\label{sec:exceptional}
\begin{figure}
\begin{equation*}
\mathcal{A}_{E_6} = 
\left(\begin{array}{cccccc}
2 & -1 & 0 & 0 & 0 & 0 \\
-1 & 2 & -1 & 0 & 0 & 0 \\
0 & -1 & 2 & -1 & 0 & -1 \\
0 & 0 & -1 & 2 & -1 & 0 \\
0 & 0 & 0 & -1 & 2 & 0 \\
0 & 0 & -1 & 0 & 0 & 2
\end{array}\right),
\qquad
\mathcal{A}_{E_7} = 
\left(\begin{array}{ccccccc}
2 & -1 & 0 & 0 & 0 & 0 & 0 \\
-1 & 2 & -1 & 0 & 0 & 0 & 0 \\
0 & -1 & 2 & -1 & 0 & 0 & 0 \\
0 & 0 & -1 & 2 & -1 & 0 & -1 \\
0 & 0 & 0 & -1 & 2 & -1 & 0 \\
0 & 0 & 0 & 0 & -1 & 2 & 0 \\
0 & 0 & 0 & -1 & 0 & 0 & 2
\end{array}\right),
\end{equation*}
\begin{equation*}
\mathcal{A}_{E_8} = 
\left(\begin{array}{cccccccc}
2 & -1 & 0 & 0 & 0 & 0 & 0 & 0 \\
-1 & 2 & -1 & 0 & 0 & 0 & 0 & 0 \\
0 & -1 & 2 & -1 & 0 & 0 & 0 & 0 \\
0 & 0 & -1 & 2 & -1 & 0 & 0 & 0 \\
0 & 0 & 0 & -1 & 2 & -1 & 0 & -1 \\
0 & 0 & 0 & 0 & -1 & 2 & -1 & 0 \\
0 & 0 & 0 & 0 & 0 & -1 & 2 & 0 \\
0 & 0 & 0 & 0 & -1 & 0 & 0 & 2
\end{array}\right)
\end{equation*}
\caption{Cartan matrices of the exceptional algebras $E_r$, $r=6,7,8$.}
\label{fig:cartanE}
\end{figure}

In this Section we study exceptional S-fold $\NN=3$ SCFTs \cite{Garcia-Etxebarria:2016erx}. We apply the techniques spelled out in the previous Sections to compute the charge lattice of these theories, the order of the 1-form symmetry group and we determine when such SCFTs can be built as discrete gauging of a parent theory. The exceptional S-fold setup of \cite{Garcia-Etxebarria:2016erx}, briefly reviewed below, engineers a set of SCFTs labelled by an algebra of type $D_n$ or $E_r$, $r=6,7,8$, and the order of the S-fold $k=3,4,6$. The analysis can be generalized by including a suitable outer automorphism \cite{Kaidi:2022lyo}. For simplicity in this paper we focus on the exceptional algebras $E_{6,7,8}$ with $k=3,4,6$ and without outer automorphism twists. Extending our methods to the full set of exceptional S-fold SCFTs should not present any major technical or conceptual difficulty, but we leave this task to future work. 
We are thus interested in 9 theories labelled by $k=3,4,6$ and by the exceptional algebra $E_r$. We find compelling arguments that suggest that all but one of these theories do not admit a well defined charge lattice and are discrete gauging of free theories. In particular the only theory that, given our current understanding, is a proper interacting $\NN=3$ SCFTs is the $G_{31}$ theory that can be engineered as the $\sfold(E_8,4)$. Our results are summarized in Table \ref{tab:results}.

In \cite{Garcia-Etxebarria:2015wns} the authors presented an alternative M-theory construction of regular S-fold theories dual to the Type IIB setup described above. This allowed them to generalize the S-fold construction of \cite{Garcia-Etxebarria:2015wns} to a wider class of theories parametrize by an ADE algebra and the order of the S-fold projection $k=3,4,6$. In this classification the regular S-folds are associated to the $A_n$ algebras, while the theories associated to the $D_n$ and $E_n$ algebras are new $\NN=3$ SCFTs that, up to now, have no known geometric construction in Type IIB.

Let us briefly review the results of  \cite{Garcia-Etxebarria:2015wns}.
The S-fold projection involves an element of the S-duality group as well as an element of the R-symmetry $SO(6)$ of $\NN=4$ $SU(N)$ $SYM$, which is a rotation transverse to the D3 branes in the Type IIB setup. By compactifying two directions $T^2_E$ transverse to the D3-brane stack we generally break the R-symmetry to $SO(4)\times \zz_2$, although for particular values of the complex structure of the torus the R-symmetry enhances to $SO(4)\times \zz_4$ for $\tau_E=i$ or $SO(4)\times \zz_6$ for  $\tau_E=e^{2\pi i /3}$. These subgroups of the original $SO(6)$ R-symmetry are enough to perform the S-fold projection.

Now we may T-dualize along one compact transverse direction, giving a Type IIA setup, and then uplift to M-theory. By carefully tracking the action of the various symmetries along these manipulations, it was shown that the regular S-fold setup is dual to M-theory on $\rr^{1,3} \times ( S_M^1 \times S_T^1 \times S_E^1 \times \cc^2)/\zz_k$, with a stack of $N$ M5-branes along  $\rr^{1,3} \times S_M^1 \times S_T^1$. The radii of the various circles are related by:
\begin{equation}
R_M = R
\qquad 
R_T = \text{Im}(\tau) R
\qquad 
R_E = \frac{1}{\text{Im}(\tau) R^2}
\end{equation}
And the $\zz_k$ quotient act as a rotation on $\cc^2$ and on the torus $S_M^1 \times S_T^1$, as well as a non-geometric quotient on $S_M^1 \times S_T^1 \times S_E^1$ fixing the $\rho$ parameter of this torus to be order 1:
\begin{equation}
\rho = \int_{T^3} C+ i \sqrt{\text{det} G}
\end{equation}
We now have an S-fold construction that involves a stack of $N$ M5-branes. Famously, on flat spacetime, this stack engineers the $(2,0)$ 6d theory of type $A_N$ once the center of mass motion is decoupled. It is natural to ask wether it is possible to generalize this setup to the other  $(2,0)$ 6d theories, namely the type $D$ and type $E$ theories.
In \cite{Garcia-Etxebarria:2015wns} it was shown that such a construction is possible and involves a non-geometric setup, meaning that there is no duality frame where the system is described by string theory in a geometric background. By contrast, the regular S-fold setup is dual to F-theory on a geometric terminal singularity.

In this paper we will study exceptional S-fold theories as a particular projection of the corresponding $\NN=4$ SYM theories obtained by compactifying the $(2,0)$ theory on a torus. Indeed both the R-symmetry and the S-duality involved in the S-fold quotient are present in the 4d theory, allowing us to understand some properties of the S-fold theories directly in 4d. There are some subtleties in this approach given by the fact that quantities of interest, for example the moduli space and the charge lattice, are only defined up to Weyl transformations of the gauge algebra, as explained in \cite{Kaidi:2022lyo}. We expand upon this approach in the rest of the paper while we refer the reader to the original literature \cite{Garcia-Etxebarria:2015wns,Kumar:1996zx} for the M-theory construction of exceptional S-fold theories.

\subsection{S-folds from the $(2,0)$ $E_6$ theory}
\label{sec:E6}
The six-dimensional $(2,0)$ theory of type $E_6$ on torus $T^2\times \mathbb{R}^4$ engineers $\NN=4$ SYM with gauge algebra $E_6$ in the 4d limit. When this compactification is complemented with the S-fold projection spelled out above one obtains the exceptional S-fold theories of interest. The strategy we adopt, introduced in full generality in Section \ref{sec:Sfolds}, is to compute the effect of the S-fold projection directly on the four-dimensional charge lattice. This approach allows us to compute the charge lattice of the $\NN=3$ S-fold theories from the charge lattice of the $\NN=4$ SYM with gauge algebra $E_6$. The analysis parallels the one in \cite{Kaidi:2022lyo}, where the moduli space of exceptional S-fold theories was computed as a subquotient of the moduli space of the $\NN=4$ SYM parent theory.

The charge lattice of $\NN=4$ $E_6$ SYM is spanned by the W-bosons, which are valued in the root lattice $\Delta$ of $E_6$ and by the magnetic monopoles, which are valued in coroot lattice $\Delta^{\vee}$. Choose a basis for the root and coroot lattices given by a set of simple roots and the corresponding coroots respectively. In this basis the metric on the root lattice is given by the Cartan matrix $\mathcal{A}_{E_6}$ of $E_6$, see Figure \ref{fig:cartanE}, and the roots are represented by integer vectors with length $\sqrt{2}$. The simple roots are represented by vectors with one entry equals to 1 and the other entries equal to 0.
A charge $\tilde{Q}$ in the charge lattice $\Gamma = \Delta \otimes \Delta^{\vee}$ is represented by an integer twelve-dimensional vectors, where the first six entries are electric charges and the last six entries are magnetic charges:
\begin{equation}	\label{eq:basis_vec_1}
\tilde{Q} = \left( e_1,e_2,\dots,e_6, m_1, m_2, \dots, m_6
\right)
\end{equation}
The Dirac pairing between two charge $\tilde{Q}$ and $\tilde{P}$ is given by $\tilde{Q} \cdot J_{E_6} \cdot \tilde{P}^{T}$ where the Dirac pairing $J_{E_6}$ is given by \cite{Argyres:2022kon}:
\begin{equation}	\label{eq:DiracE6}
J_{E_6} = 
\left(
\begin{array} {cc}
0 & \left( \mathcal{A}_{E_6} \right)^{T}
\\
 -\mathcal{A}_{E_6} & 0
\end{array}
\right)
\end{equation}

In the following it will be more convenient to write the charges in a basis where we alternate electric charges and magnetic charge, namely:
\begin{equation}	\label{eq:basis_vec_2}
Q = \left( e_1, m_1, e_2, m_2, \dots, e_6,m_6
\right)
\end{equation}
We will distinguish the charges in the two basis by using tildes for vectors in the first basis \eqref{eq:basis_vec_1} and symbols without tildes in the second basis \eqref{eq:basis_vec_2}. 

The Weyl group of $E_6$ is generated by the reflections along the simple roots, we denote the reflection along the $i$-th simple root as $s_i$. A useful element of the Weyl group is the Coxeter element $c_{E_6}$, defined as:
\begin{equation}
c_{E_6} = s_1 \cdot s_2 \cdot s_3 \cdot s_4 \cdot s_5 \cdot s_6 
\end{equation}
which has order equal to the Coxeter number $h_{E_6} = 12$:
\begin{equation}
\left(c_{E_6} \right)^{12} = \text{Id}
\end{equation}
The eigenvalues of the Coxeter element are $\lambda_i = e^{2\pi i / (m_i-1)}$ where $m_i$, $i=1\dots,6$ are the degrees of the invariants of $E_6$, tabulated in \ref{fig:degE}. 
In the basis given by the simple roots the Coxeter element is represented by the matrix:

\begin{equation}
\settowidth\mylen{$-1$}
c_{E_6}=
\left(\begin{array}{*{6}{wc{\mylen}}}
\- 0 &\; 0 & 1 & 0 & -1 & -1 \\
1 & 0 & 1 & 0 & -1 & -1 \\
0 & 1 & 1 & 0 & -1 & -1 \\
0 & 0 & 1 & 0 & -1 & 0 \\
0 & 0 & 0 & 1 & -1 & 0 \\
0 & 0 & 1 & 0 & 0 & -1
\end{array}\right)
\end{equation}

\begin{table}
\centering
\begin{tabular}{|c|c|c|}
\hline
$ \mathfrak{g} $& degrees & codegrees
 \\ \hline
 $E_6$ & 2,5,6,8,9,12 & 0,3,4,6,7,10
  \\ \hline
 $E_7$  & 2,6,8,10,12,14,18 & 0,4,6,8,10,12,16
  \\ \hline
 $E_8$ & 2,8,12,14,18,20,24,30 & 0,6,10,12,16,18,22,28
 \\ \hline
\end{tabular}
\caption{Degrees and codegrees of the exceptional algebras $E_r$.}
\label{fig:degE}
\end{table}

\begin{table}
\centering
\begin{tabular}{|c|c|c|c|c|}
\hline
\multicolumn{5}{|c|}{$E_6$}\\
\hline
$k$ & $w$ & $r$ & ECCRG & $\Delta_i $\\
\hline
3 & $\left(c_{E_6} \right)^4$ & 3 & $G_{25}$ & $\{6,9,12\}$ \\
\hline
4 & $\left(c_{E_6} \right)^3$ & 2 & $G_8$ & $\{8,12\} $\\
\hline
6 & $\left(c_{E_6} \right)^2$ & 2 & $G_5$ & $\{6,12\} $\\
\hline
\end{tabular}
\caption{Elements $w\in \mathcal{W} \left[E_6\right]$ that characterize the S-fold actions for the S-fold theories of type $E_6$ through \eqref{eq:CB_E6} and \eqref{eq:states_E6}. Here $c_{E_6}$ is the Coxeter element of $E_6$, $r$ is the rank of the S-fold theory and in the fourth and fifth columns the exceptional complex crystallographic reflection group (ECCRG) associated to the S-fold theories and its degrees $\Delta_i$ are reproduced.}
\label{fig:wE6}
\end{table}

Consider now the exceptional S-fold setup that engineers an $\NN=3$ SCFT in four dimensions. 
In Section \ref{sec:Sfolds} we studied the S-fold projection along the lines of \cite{Kaidi:2022lyo} and discussed how the rank, CB , charge lattice and associated Dirac pairing can be computed directly from the $\NN=4$ parent theory, in this case $\NN=4$ $E_6$ SYM. Here we summarize the main results for ease of readibility.
The CB  of the $\NN=3$ S-fold theory is given by the solutions to:
\begin{equation}	\label{eq:CB_E6}
w\cdot \phi_{\mathcal{C}} = e^{2\pi i/k} \phi_{\mathcal{C}}
\end{equation}
where $\phi_{\mathcal{C}}$ are elements of the CB  of $E_6$ $\NN=4$ SYM. 
The element $w\in \mathcal{W} \left[ E_6 \right]$ encodes 
the projection induced by the S-fold on the CB  and on the charge lattice of the $E_6$ $\NN=4$ theory.
The rank $r$ of the $\NN=3$ theory is given by the complex dimension of the eigenspace associated to the eigenvalue $e^{2\pi i/k}$ of $w$ and
we choose $w$ such that the $\NN=3$ has maximum rank, following \cite{Kaidi:2022lyo}.
The degrees of basic CB  invariants is then given by the degrees of invariants of $E_6$ that are divisible by $k$ and the $\NN=3$ CB  itself is $\cc^r/G$ with $G$ the complex reflection group with the correct degrees, see Table \ref{fig:wE6}.

The charged states $\overline{|\alpha;(p,q)\rangle}$ of the $\NN=3$ theory are given by:
\begin{equation}	\label{eq:states_E6}
\overline{|\alpha;(p,q)\rangle} = 
\frac{1}{\sqrt{k}} \sum_{t=0}^{k-1}\left|w^{t } \cdot \alpha ;\left(\rho_k\right)^t \cdot(p, q)\right\rangle
\end{equation}
where $|\beta; (p,q)\rangle$ is a $(p,q)$-dyonic states of  $E_6$ $\NN=4$ SYM associated to the root $\beta$ of $E_6$. The electromagnetic charge of a state $\overline{|\alpha;(p,q)\rangle}$ is given by;
\begin{equation}	\label{eq:Q_E6}
Q\left[\overline{|\alpha;(p,q)\rangle} \right] = 
\frac{1}{\sqrt{k}} \sum_{t=0}^{k-1}
(w \otimes \rho_k )^t \cdot
Q\left[|\alpha; (p,q)\rangle\right]
\end{equation}
where $Q\left[|\alpha; (p,q)\rangle\right]$ is the electromagnetic charge of the corresponding state of  $E_6$ $\NN=4$ SYM, expressed as in \eqref{eq:basis_vec_2}. As an example the W-boson associated to the first root $\alpha_1$ of $E_6$ has charge $Q\left[|\alpha_1; (1,0)\rangle\right]= (1,0;0,0; \dots)$ while the magnetic monopole associated to the first coroot has charge $Q\left[|\alpha_1; (0,1)\rangle\right]= (0,1;0,0; \dots)$.

One can consider more general states that are invariant under the S-fold action, see for example \eqref{eq:CH_state}. In the case of regular S-folds some of these states appear in the presence of discrete torsion and correspond to strings stretched between the S-fold and a D3 brane. We checked that in the case of exceptional S-folds the states \eqref{eq:CH_state} can never be included consistently because they break the Dirac quantization condition, therefore in the remainder of this paper we will only mention them briefly.

Finally, the Dirac pairing defined on the charge lattice of the $\NN=3$ theory is obtained as a restriction of the Dirac pairing of $E_6$ $\NN=4$ SYM \eqref{eq:DiracE6}. Explicitly the Dirac pairing between two states of the S-fold theories with charges $q_i$ and $q_j$ is given by:
\begin{equation}	\label{eq:J_E6}
\left\langle q_i, q_j\right\rangle=q_i \cdot J_{E_6} \cdot q_j^T
\end{equation}
Notice that it is not guaranteed that $\left\langle q_i, q_j\right\rangle$ gives an integer result and one should check case by case that the Dirac pairing between any two charges of the S-fold theories is integer. In the following we do not consider any charge lattice where the Dirac pairing can take fractional values.

\subsection{The $k=6$ S-fold: $G_5$}
\label{sec:G5}
The first exceptional S-fold theory that we consider is obtained as a $\zz_6$ S-fold compactification of the $(2,0)$ six-dimensional $E_6$ theory to four dimension. 
The compactification preserves $\NN=3$ supersymmetry in four dimension and involves an S-duality transformation $\rho_6\in SL(2,\zz)$ and an R-symmetry twist.
The CB  is given by
the solutions to \eqref{eq:CB_E6} with $k=6$, namely:
\begin{equation}	\label{eq:CB_E6_G5}
w\cdot \phi_{\mathcal{C}} = e^{\pi i/3} \phi_{\mathcal{C}}
\end{equation}
There are 2 invariants of $E_6$ whose degrees are divisible by 6, namely the invariants with degrees 6 and 12, and therefore we expect that the $\NN=3$ theory has rank $r=2$. We choose an element $w\in \mathcal{W} \left[ E_6 \right]$ which has a two-dimensional eigenspace associated to the eigenvalue $e^{\pi i/3}$:
\begin{equation}
	w = \left( c_{E_6}\right)^2
\end{equation}
which is the basis given by the simple roots is represented by the matrix:
\begin{equation}
\settowidth\mylen{$-1$}
w = 
\left(\begin{array}{*{6}{wc{\mylen}}}
0 & 1 & 0 & -1 & 0 & 0 \\
0 & 1 & 1 & -1 & -1 & -1 \\
1 & 1 & 1 & -1 & -1 & -1 \\
0 & 1 & 1 & -1 & 0 & -1 \\
0 & 0 & 1 & -1 & 0 & 0 \\
0 & 1 & 0 & 0 & -1 & 0
\end{array}\right)
\end{equation}
Then the CB  of the $\NN=3$ theory, given by the solutions to \eqref{eq:CB_E6_G5}, is $\cc^2/G_5$ where $G_5$ is the CCRG with degrees 6 and 12. Similarly the charge lattice of the $\NN=3$ theory can be obtained from the charge lattice of the $\NN=4$ $E_6$ SYM. Given a state of $\NN=4$ $E_6$ SYM associated to the root $\alpha$ with electric and magnetic charges $(p,q)$ one can build a state $\overline{|\alpha,(p, q)\rangle}$ that is invariant under the S-fold action:
\begin{equation}	\label{eq:N3states}
\overline{|\alpha,(p, q)\rangle}=\frac{1}{\sqrt{6}} \sum_{t=0}^{5}\left|w^{t} \cdot \alpha ;\left(\rho_6\right)^t \cdot(p, q)\right\rangle
\end{equation}
Consider the six states $\overline{|\alpha_i,(1, 0)\rangle}$ obtained with this projection from the $W$-bosons associated to the simple roots $\alpha_i$, $i=1,\dots,6$ of $E_6$. The electromagnetic charges $q_i$ of these states can be computed using \eqref{eq:Q_E6}:
\begin{equation}
\settowidth\mylen{$-1$}
\begin{array}{cc*{12}{wc{\mylen}}}
q_1 =&\displaystyle \frac{1}{\sqrt{6}} &(2,& -1,& 1,& -2,& 2,& -4,& 0,& -3,& 1,& -2,& 0, &0)
\\
q_2=&\displaystyle\frac{1}{\sqrt{6}} &(1,& -2,& 3,& -3,& 2, &-4,& 1,& -2,& -1,& -1,& 2,& -4)
\\
q_3=&\displaystyle\frac{1}{\sqrt{6}} &(1,& 1,& 2,& -1,& 4,& -2,& 3,& -3,& 2,& -1,& 0,& 0)
\\
q_4=&\displaystyle\frac{1}{\sqrt{6}} & (-2,& 1,& -1,& 2,& -2,& 4,& 0,& 3,& -1,& 2,& 0,& 0)
\\
q_5=&\displaystyle\frac{1}{\sqrt{6}} & (-1,& 2,& -3,& 3,& -2,& 4,& -1,& 2,& 1,& 1,& -2,& 4)
\\
q_6=&\displaystyle\frac{1}{\sqrt{6}} &(0,& 0,& -2,& 4,& -2,& 4,& -2,& 4,& 0,& 0,& 2,& 2)
\end{array}
\end{equation}
Notice that $q_1=-q_4$ and $q_2 = -q_5$, therefore these charges span a four-dimensional lattice $\Gamma$:
\begin{equation}	\label{eq:GammaG5}
\Gamma = \text{Span}_{\mathbb{Z}} \left\{  q_1, q_2, q_3, q_6	\right\}
\end{equation}
 The charges of states obtained from $W$-bosons associated to other roots of $E_6$ are included in $\Gamma$ because the other roots are linear integer combinations of the simple roots and \eqref{eq:Q_E6} is linear in the charges. One can also check that the charges of states $\overline{|\alpha_i,(0, 1)\rangle}$  obtained from monopoles of $E_6$ are included in $\Gamma$ as well, therefore by linearity $\Gamma$ includes the charges of all the states \eqref{eq:N3states}. 
One may also consider the more general states \eqref{eq:CH_state}.
We checked explicitly that including some or all of these states either leaves $\Gamma$ unchanged or produces fractional Dirac pairing between the states, which is inconsistent.
Then	 \eqref{eq:GammaG5} is the candidate for the charge lattice $\Gamma$ of the $\NN=3$ $G_5$ exceptional S-fold theory. In the remainder of this section we will show that $\Gamma$ is actually incompatible with a consistent CB  stratification, and we will argue that the low energy field theory is given by a discrete gauging of free $U(1)^2$ $\NN=4$ gauge theory.

Having computed a candidate $\Gamma$ for the charge lattice of the $G_5$ theory we now study the Dirac pairing defined on this lattice and the sublattices of states that become massless on some CB  singularity.
The Dirac pairing between two states with charges $q_i$ and $q_j$ is given by:
\begin{equation}
\langle q_i, q_j \rangle = q_i \cdot J_{E_6} \cdot q_j^{T}
\end{equation}
where $J_{E_6}$ is the Dirac pairing of the $\NN=4$ theory \eqref{eq:DiracE6}. In the basis of $\Gamma$ given by $q_1, q_2, q_3$ and $q_6$ the Dirac pairing $J_{G_5}$ of the $\NN=3$ theory is represented by the matrix:
\begin{equation}
	J_{G_5} = 
	\left(\begin{array}{cccc}
0 & -1 & 3 & -4 \\
1 & 0 & -1 & 4 \\
-3 & 1 & 0 & -2 \\
4 & -4 & 2 & 0
\end{array}\right)
\end{equation}
If $\Gamma$ is the charge lattice of the $G_5$ theory then the order of the 1-form symmetry group is given by the absolute value of the Pfaffian of $J_{G_5}$:
\begin{equation}
\left|G^{(1)}_{G_5} \right| = \left|\pf[J_{G_5}] \right| = 6
\end{equation}

Let us consider the states becoming massless on some codimension-1 singularity on the CB . We can parametrize the CB  of the $\NN=4$ $E_6$ SYM with six complex scalars $\phi_i$, $i=1,\dots,6$, with identifications given by the Weyl group of $E_6$. The CB  $\mathcal{C}_{G_5}$ of the $\NN=3$ $G_5$ theory is given by the eigenspace of $w$ with eigenvalue $e^{\pi i/3}$ and can be parametrized as follows as en embedding in $\mathcal{C}_{E_6}$:

\begin{equation}	\label{eq:CB_G5}
\begin{split}
\mathcal{C}_{G_5} =&  \left\{
v_3  \phi_3 + v_4 \phi_4 , \;\; \phi_3, \phi_4 \in \cc
\right\} \cap \mathcal{C}_{E_6}
\\
v_3 = &  \left( 1,e^{1\pi/3}, 1,0,-e^{2\pi i/3} , \sqrt{3} e^{i\pi/6} \right)
\\
v_4 = &  \left( -1, e^{-i\pi/3},0,1,e^{2\pi i/3} ,2e^{4\pi i/3}	\right)
\end{split}
\end{equation}
The codimension-1 singularities of $\mathcal{C}_{G_5}$ correspond to fixed points under the reflection of $G_5$ acting on this slice. As discussed in Section \ref{sec:moduli_space} can be obtained as the intersections of the codimension-1 singularities of the $E_6$ $\NN=4$ SYM with the slice \eqref{eq:CB_G5}. As an example consider the singularity $\mathcal{H}_{s_1}^{E_6}$ of $\mathcal{C}_{E_6}$ corresponding to the fixed locus under $s_1$, the reflection along the first simple root of $E_6$, which is the 5-dimensional hyperplane:
\begin{equation}
\mathcal{H}_{s_1}^{E_6} = \left\{\left(
\phi_1, \; 2\phi_1, \; \phi_3, \; \phi_4, \; \phi_5, \; \phi_6 \right),
\phi_i \in \cc
\right\}
\end{equation}
The intersection of $\mathcal{H}_{s_1}^{E_6}$ with the slice $\mathcal{C}_{G_5}$ gives a codimension-1 singularity $\mathcal{H}_{s_1}^{G_5} $ of the CB  of the $G_5$ theory:
\begin{equation}
\begin{split}
\mathcal{H}_{s_1}^{G_5} &= \mathcal{H}_{s_1}^{E_6} \cap \mathcal{C}_{G_5}
\\
&=
\left(
1,2 ,\frac{1}{2} \left(5-i \sqrt{3}\right),\frac{1}{2} \left(3-i \sqrt{3}\right),-(-1)^{2/3},\frac{1}{2} \left(3-i \sqrt{3}\right) 
\right)\phi _3
\end{split}
\end{equation}
The states that can become massless on $\mathcal{H}_{s_1}^{G_5}$ are those whose central charge $Z$ vanish identically on $\mathcal{H}_{s_1}^{G_5}$.
 On a generic point $\mathbf{\phi}$ of the CB  of $E_6$ $\NN=4$ SYM the central charge $Z$ of a state with charge $q$ is given by:
\begin{equation}
Z[q] = \sum_{i,j=1}^{6} \phi_i  \left(A_{E_6}\right)_{ij} (e_j + \tau m_j)
\end{equation}
The central charges $Z[q_1]$ and $Z[q_3]$ of $\overline{|\alpha_1,(1, 0)\rangle}$ and $\overline{|\alpha_3,(1, 0)\rangle}$ identically vanish on the singularity $\mathcal{H}_{s_1}^{G_5}$, therefore the corresponding BPS states become massless on this singularity. One can also check that the sublattice $\Gamma^{\mathcal{H}_{s_1}^{G_5}}$ of charges of states that become massless on this singularity $\mathcal{H}_{s_1}^{G_5}$ is generated by $q_1$ and $q_3$. The lattice $\Gamma^{\mathcal{H}_{s_1}^{G_5}}$ should correspond to the charge lattice of the rank-1 CFT supported on the singularity $\mathcal{H}_{s_1}^{G_5}$. 
The Dirac pairing restricted to the sublattice $\Gamma^{\mathcal{H}_{s_1}^{G_5}}$, which we denote as $J^{\mathcal{H}_{s_1}^{G_5}}$, has Pfaffin given by:
\begin{equation}
\left|\pf[J^{\mathcal{H}^{G_5}_{s_1}}] \right| = \left| \langle q_1, q_3 \rangle\right| = 3
\end{equation}
Which should be equal to the order of the 1-form symmetry group of the rank-1 theory supported on the singularity  $\mathcal{H}_{s_1}^{G_5}$. Then the theory on this singularity would be a rank-1 $\NN \geq 2$ SCFT with a 1-form symmetry group of order 3. All the rank-1 theories with $\NN=2$ or higher supersymmetry have been classified, and such a theory does not exist. In particular the maximum order of the 1-form symmetry group for a rank-1 $\NN=2$ SCFT is 2.
We conclude that the theory living on this singularity of CB  is not a CFT, but rather a discrete gauging of free $U(1)$ $\NN=4$ Maxwell theory, which is the only other possibility\footnote{Remember that the exceptional S-fold theory are maximally strongly coupled, therefore the theories living on the singularities of the moduli space can not be IR-free theories.}.
In particular this implies that there are no states becoming massless on the singularity, therefore the states with charges lying on the sublattice  $\Gamma^{\mathcal{H}_{s_1}^{G_5}}$ are not BPS.

As another example, consider the singularity corresponding to the reflection $s_6$ along the sixth root of $E_6$. The locus of the singularity $\mathcal{H}_{s_6}^{G_5}$ can be parametrized as:
\begin{equation}
\mathcal{H}_{s_6}^{G_5}=
\left(
1,\frac{1}{6} \left(9-i \sqrt{3}\right),\left(2-\frac{2 i}{\sqrt{3}}\right),\left(1-\frac{2 i}{\sqrt{3}}\right),-(-1)^{2/3} ,\left(1-\frac{i}{\sqrt{3}}\right)
\right) \phi _3
\end{equation}
and the sublattice of charges becoming massless on this singularity is spanned by $q_6$ and $(q_2+q_3-q_1)$. The Dirac pairing restricted to this sublattice has Pfaffian equal to:
\begin{equation}
\left|\pf[J^{\mathcal{H}^{G_5}_{s_6}}] \right| = \left| \langle q_6, q_2+q_3-q_1 \rangle\right| = 6
\end{equation}
Then the rank-1 CFT on this singularity should have a 1-form symmetry group of order 6. As was the case for the previous singularity, such a CFT does not exist, and the theory on this singularity must be a discrete gauging of free $U(1)$ $\NN=4$ Maxwell theory. We conclude that there are no states becoming massless on this singularity.

One can perform similar computations on all the singularities of the CB  of the $G_5$ $\NN=3$ theory. It turns out that all the codimension-1 singularities are equivalent, up to $G_5$ transformations, either to $\mathcal{H}_{s_1}^{G_5}$ or to $\mathcal{H}_{s_6}^{G_5}$. It follows that the rank-1 theories supported on every codimension-1 singularities of the CB  $\mathcal{C}_{G_5}$ are discrete gaugings of free $U(1)$ $\NN=4$ Maxwell theory. Then there are no charged states becoming massless an any codimension-1 singularity, and the $G_5$ theory itself must be a discrete gauging of a free theory, namely free $U(1)^2$ $\NN=4$ gauge theory. 
Indeed, if any charged state with charge $q$ become massless at the origin of the CB , which is a codimension-2 singularity, then it satisfies the BPS bound and is massless whenever its central charge vanishes, namely on the codimension-1 hypersurface identified by $Z[q]=0$. As we just discussed there are no charged states that become massless on any codimension-1 singularities, therefore there are no massless charged states on any point of the CB , including the origin. In section \ref{sec:bound} we give additional evidence for this claim and show that it is in fact  impossible to define a consistent charge lattice on a CB  with geometry $\cc^2/G_5$.

\subsection{The $k=4$ S-fold: $G_8$}
\label{sec:G8}
In this Section we consider the exceptional S-fold SCFT obtained as a $k=4$ S-fold of the $E_6$ $(2,0)$ six-dimensional theory, called the $G_8$ SCFT. We find that the charge lattice is not consistent with the stratification proposed in \cite{Kaidi:2022lyo}. In more detail, our analysis suggests that the theory supported on condimension-1 singularities in the CB  is the $\NN=3$ preserving $\zz_4$ gauging of $SU(2)$ $\NN=4$ SYM, while the constraints from the central charge formulae are compatible with this theory being the rank-1 $S_{4,4}$-fold SCFT. Therefore we claim that the $G_8$ theory is a discrete gauging of free $U(1)^2$ $\NN=4$ gauge theory.\\

The CB  of the $G_8$ theory is given by the solutions of \eqref{eq:CB_E6} with $w$ an element of the Weyl group of $E_6$:
\begin{equation}
w = \left(c_{E_6}\right)^3
\end{equation} 
which satisfies $w^4=\text{Id}$ and has a two-dimensional eigenspace with eigenvalue $e^{\pi i /2}$. The $\NN=3$ theory then has rank $r=2$ and the degrees of invariants on the CB  are given by the degrees of $E_6$ that are divisible by 4, namely $\Delta_i = {8,12}$. The CB  is given by $\cc^2/G_{8}$ where $G_{8}$ is the exceptional complex reflection group with the correct degrees of invariants.

The states that are invariant under the S-fold action can be computed using \eqref{eq:states_E6}, \eqref{eq:Q_E6}. In particular the states obtained from the W-bosons corresponding to the simple roots of $E_6$ have charges:
\begin{equation}
\settowidth\mylen{$-1$}
\begin{array}{c*{12}{wc{\mylen}}}
 \mathrm{q}_1=\displaystyle \frac{1}{\sqrt{2}}&(1,&-1 ,& 0,&-2 ,& 0,&-2 ,&-1,&-1 ,& 0,&0 ,& 0,&-2) \\
 \mathrm{q}_2=\displaystyle \frac{1}{\sqrt{2}}&(0,&0 ,& 1,&-1 ,& 0,&-2 ,& 0,&-2 ,&-1,&-1 ,& 0,&0) \\
 \mathrm{q}_3=\displaystyle \frac{1}{\sqrt{2}}&(1,&-1 ,& 1,&-1 ,& 2,&-2 ,& 1,&-1 ,& 1,&-1 ,& 0,&-2) \\
 \mathrm{q}_4=\displaystyle \frac{1}{\sqrt{2}}&(-1,&1 ,& 0,&2 ,& 0,&2 ,& 1,&1 ,& 0,&0 ,& 0,&2) \\
 \mathrm{q}_5=\displaystyle \frac{1}{\sqrt{2}}&(0,&0 ,&-1,&1 ,& 0,&2 ,& 0,&2 ,& 1,&1 ,& 0,&0) \\
 \mathrm{q}_6=\displaystyle \frac{1}{\sqrt{2}}&(0,&2 ,& 0,&2 ,& 0,&4 ,& 0,&2 ,& 0,&2 ,& 2,&2)
\end{array}
\end{equation}
Notice that $q_4 = -q_1$ and $q_5=-q_2$, therefore these charges span a four-dimensional lattice. By computing the charges of the states obtained from the magnetic monopoles of the $E_6$ $\NN=4$ theory and by linearity argument one shows that the candidate $\Gamma$ for charge lattice $\Gamma$ of the $G_8$ theory is:
\begin{equation}	\label{eq:chargelattice_G8}
\Gamma = \text{Span}_{\mathbb{I}} \left\{  q_1, q_2, q_3, q_6	\right\}
\end{equation}
The charge lattice in the basis $\left\{  q_1, q_2, q_3, q_6	\right\}$ is represented by the matrix  $J_{G_8}$:
\begin{equation}
J_{G_8}=
\left(\begin{array}{cccc}
0 & 1 & -1 & 2 \\
-1 & 0 & 1 & -2 \\
1 & -1 & 0 & 2 \\
-2 & 2 & -2 & 0
\end{array}\right)
\end{equation}
And the order of the 1-form symmetry group is given by:
\begin{equation}
\left|G_{G_8}^{(1)}\right|=\left|\operatorname{Pf}\left(J_{G_8}\right)\right|=2
\end{equation}
Next we can study the sublattices of charges of states becoming massless at codimension-1 singularities. All the codimension-1 singularities are related by $G_8$ transformations and the slices transverse to these singularities are locally $\cc/\zz_4$. Furthermore through similar computations to the ones spelled out in the previous section one finds that the charge lattice of the rank-1 theory supported on these singularities is generated by two charges $Q_1$ and $Q_2$ with $|\langle Q_1, Q_2 \rangle| = 2$. 
Then the rank-1 theory supported on the codimension-1 singularities is a $\NN \geq 2$ SCFTs with CB  $\cc/\zz_4$ and a non-trivial $\zz_2$ 1-form symmetry. The only candidate is the $\NN=3$ preserving $\zz_4$ descrete gauging of $\NN=4$ $SU(2)$ SYM \cite{Argyres:2016yzz}. This is in contradiction with the analysis of the central charge of the $G_8$ theory performed in \cite{Cecotti:2021ouq,Kaidi:2022lyo}, where the theory supported on the codimension-1 singularities was found to be the rank-1 $S_{4,4}$-fold SCFT, denoted also as $\mathcal{S}_{\varnothing, 4}^{(1)}$. Let us briefly review this analysis.

Assuming that the $G_8$ theory is not a discrete gauging, the central charges $a = c$ can be computed with the Shapere-Tachikawa formula \cite{Shapere:2008zf}:
\begin{equation}
2(2a-c) = 
\sum_{j=1}^{r} \Delta_j -\frac{r}{2} 
\end{equation}
where $\Delta_i$ are the degrees of the fundamental invariants on the CB . In the case of the $G_8$ theory we have $\{ \Delta_1, \Delta_2\} = \{8,12\}$. On the other hand the formulae of \cite{Martone:2020nsy, Martone:2021ixp} allow us to relate the central charges of the $G_8$ theory with the data of the rank-1 theories supported on the codimension-1 singularities:
\begin{equation}
12 c=2 r+h_{\mathrm{ECB}}+\sum_{i \in \mathcal{I}} \Delta_i^{\mathrm{sing}} b_i
\end{equation}
where $b_i$ is a quantity associated with the rank-1 theory supported on the codimension-1 singularities as follows:
\begin{equation}	\label{eq:b_martone}
b_i:=\frac{12 c_i-h_i-2}{\Delta_i}
\end{equation}
In our theory we have $h_{\mathrm{ECB}} = r = 2$ and the set $\mathcal{I}$ of strata consist of only one singularity with scaling dimension $ \Delta^{\mathrm{sing}} = \text{l.c.m.}(8,12) = 24$ and parameter $b$. Then, remembering that $a=c$ for any $\NN=3$ theory, one may solve for $b$ and finds:
\begin{equation}
b=\frac{9}{2}
\end{equation}
which is compatible with the rank-1 fluxfull S$_{4}$-fold SCFT. In contrast, if the theory supported on the codimension-1 singularities was a discrete gauging of $SU(2)$ $\NN=4$ SYM, we would have $b_{SU(2)}=3$.\\

We found that the charge lattice \eqref{eq:chargelattice_G8} is incompatible with analysis of the central charges performed via the stratification of the CB \footnote{Another possibility is that the Shapere-Tachikawa formula does not hold for the $G_8$ theory. In that case the $G_8$ theory could be an interacting SCFT with $12c=78$. We do not consider this possibility further in this paper and trust the Shapere-Tachikawa formula for any theory that is not a discrete gauging.}.
 Therefore we claim that charged states can not become massless on the singularities of the CB . Similar to the case of the $G_5$ theory, studied in Section \ref{sec:G5}, we thus conclude that the $G_8$ theory is not an interacting SCFT but rather a discrete gauging of free $U(1)^2$ $\NN=4$ gauge theory. In Section \ref{sec:bound} we will give additional evidence for this claim by showing that any well defined charge lattice on the CB  $\cc^2/G_8$ is only compatible with having (a discrete gauging of) $SU(2)$ $\NN=4$ SYM supported on the codimension-1 singularities.

\subsection{The $k=3$ S-fold: $G_{25}$}
In this section we study the theory obtained with a $k=3$ exceptional S-fold from the $E_6$ $(2,0)$ six-dimensional theory, denoted as the $G_{25}$ theory.
By similar argument to the ones spelled out in the previous cases we find that this theory is a discrete gauging of $U(1)^3$ $\NN=4$ gauge theory. 
In particular the charge is incompatible with any choice of rank-1 SCFTs on the codimension-1 singularities of the CB .

 The CB  and charge lattice can be found respectively with \eqref{eq:CB_E6} and \eqref{eq:Q_E6} with:
 \begin{equation}
w=\left(c_{E_6}\right)^4
\end{equation}
which satisfies $w^3=\text{Id}$ and has a three-dimensional eigenspace with eigenvalue $e^{2\pi i /3}$. The $\NN=3$ theory then has rank $r=3$ and the degrees of invariants on the CB  are given by the degrees of $E_6$ that are divisible by 3, namely $\Delta_i = {6,9,12}$. The CB  is given by $\cc^3/G_{25}$ where $G_{25}$ is the exceptional complex reflection group with the correct degrees of invariants.  

The lattice of electromagnetic charges associated to the rank-1 theory supported on the codimension-1 singularities can be computed with the techniques spelled out in the previous Sections. 
The result is that these lattices are generated by two charges $Q_1$ and $Q_2$ with $\langle Q_1, Q_2 \rangle=3$, indicating that the rank-1 theory on these singularity should have a 1-form symmetry group of order 3. This is not possible and we conclude that the theory on the codimension-1 singularity is a discrete gauging of $\NN=4$ Maxwell theory. Therefore the $G_{25}$ theory itself must be a discrete gauging of free $U(1)^3$ $\NN=4$ gauge theory, because charged states can not become massless anywhere on the CB . In Section \ref{sec:bound} we give additional evidence for this claim and show that it is impossible to define a consistent charge lattice on a CB  $\cc^3/G_{25}$.

\subsection{S-folds from the $(2,0)$ $E_7$ theory}	
\label{sec:E7}

\begin{table}
\centering
\begin{tabular}{|c|c|c|c|c|}
\hline
\multicolumn{5}{|c|}{$E_7$}\\
\hline
$k$ & $w$ & $r$ & ECCRG & $\Delta_i $\\
\hline
3 & $\left(c_{E_7} \right)^6$ & 3 & $G_{26}$ & $\{6,12, 18\}$ \\
\hline
4 & $\left(c_{E_6} \right)^3$ & 2 & $G_8$ & $\{8,12\} $\\
\hline
6 & $\left(c_{E_7} \right)^3$ & 3 & $G_{26}$ & $\{6,12, 18\} $\\
\hline
\end{tabular}
\caption{Elements $w\in \mathcal{W} \left[E_7\right]$ that characterize the S-fold actions for the S-fold theories of type $E_7$ through \eqref{eq:CB_E6} and \eqref{eq:states_E6}. Here $c_{E_7}$ and $c_{E_6}$ are the Coxeter element of $E_7$ and the $E_6$ subalgebra, respectively, $r$ is the rank of the S-fold theory and in the fourth and fifth columns the exceptional complex crystallographic reflection group (ECCRG) associated to the S-fold theories and its degrees $\Delta_i$ are reproduced.}
\label{fig:wE7}
\end{table}

In this section we consider the exceptional S-fold theories obtained from the $(2,0)$ six dimensional theory of type $E_7$. All the techniques that we use were spelled out in details in Section \ref{sec:Sfolds} and were applied to the $E_6$ case in Section \ref{sec:E6}. Therefore in this section we will only provide the informations that define the S-fold projection, namely the element $w\in \mathcal{W}\left[E_7\right]$, and the final results. 
The main result is that all the exceptional S-folds SCFTs obtained from the $E_7$ theories are discrete gauging of free $U(1)^r$ $\NN=4$ gauge theory, where $r$ is the rank of the theory, see Table \ref{fig:wE7}.

We work in a basis of the algebra $E_7$ given by simple roots $\alpha_i$ such that the Cartan matrix is the one in Figure \ref{fig:cartanE}. The reflections along the simple roots are denoted as $s_i$ and the corresponding Coxeter element of $E_7$ is:
\begin{equation}
\settowidth\mylen{$-1$}
c_{E_7} = \prod_{i=1}^{7} s_i =
\left(\begin{array}{*{7}{wc{\mylen}}}
 0 & 0 & 0 & 1 & 0 & -1 & -1 \\
 1 & 0 & 0 & 1 & 0 & -1 & -1 \\
 0 & 1 & 0 & 1 & 0 & -1 & -1 \\
 0 & 0 & 1 & 1 & 0 & -1 & -1 \\
 0 & 0 & 0 & 1 & 0 & -1 & 0 \\
 0 & 0 & 0 & 0 & 1 & -1 & 0 \\
 0 & 0 & 0 & 1 & 0 & 0 & -1 \\
\end{array}
\right)
\end{equation}
which satisfies:
\begin{equation}
\left(c_{E_7}\right)^{18} = \mathbb{1}
\end{equation}
In defining the elements $w$ involved in the S-fold projections we will also use the Coxeter element of the $E_6$ subalgebra:
\begin{equation}
\settowidth\mylen{$-1$}
c_{E_6\subset E_7} =   \prod_{i=2}^{7} s_i  =
\left(\begin{array}{*{7}{wc{\mylen}}}
 1 & 0 & 0 & 0 & 0 & 0 & 0 \\
 1 & 0 & 0 & 1 & 0 & -1 & -1 \\
 0 & 1 & 0 & 1 & 0 & -1 & -1 \\
 0 & 0 & 1 & 1 & 0 & -1 & -1 \\
 0 & 0 & 0 & 1 & 0 & -1 & 0 \\
 0 & 0 & 0 & 0 & 1 & -1 & 0 \\
 0 & 0 & 0 & 1 & 0 & 0 & -1 \\
\end{array}
\right)
\end{equation}
which satisfies:
\begin{equation}
\left(c_{E_6\subset E_7}\right)^{12} = \mathbb{1}
\end{equation}

The degrees and codegrees of $E_7$ are tabulated in Table \ref{fig:degE}. Let us now consider the exceptional S-fold theories parametrized with $k=3,4,6$.

\begin{itemize}
\item{\textbf{Case $\mathbf{k=3}$, $\mathbf{G_{26}}$:} }
The CB  and charge lattice can be computed with \eqref{eq:CB_E6} and \eqref{eq:states_E6} respectively with:
\begin{equation}
w=\left(c_{E_7}\right)^{6}
\end{equation}
The theory is a rank 3 SCFT with CB  $\cc^3/G_{26}$ where $G_{26}$ is the ECCRG with degrees 6,12 and 18. There are two independent codimension-1 singularities that correspond to two rank-2 CBs with geometry $\cc^2/G_5$ and $\cc^2/G(3,1,2)$, respectively. The slice transverse to the $G_5$ singularity is $\cc/\zz_2$ while the slice transverse to the $G(3,1,2)$ singularity is $\cc/\zz_3$. One can compute the order of the 1-form symmetry groups of the rank-1 theories supported on these singularities from the charge lattice, and we find:
\begin{equation}
\zz_2 \text{ singularity: } G^{(1)} = \zz_2
\qquad
\qquad
\zz_3 \text{ singularity: } G^{(1)} = \zz_3
\end{equation}
There is no rank-1 $\NN=2$ SCFT with a $\zz_3$ 1-form symmetry, therefore we conclude that the $\zz_3$ singularity is empty and supports a discrete gauging of free $U(1)$ $\NN=4$ Maxwell theory.
Comparing with the analysis of the central charges performed in \cite{Kaidi:2022lyo} the only option is that the $\zz_2$ singularity is empty as well and therefore the $G_{26}$ theory is itself a discrete gauging of free $U(1)^3$ $\NN=4$ gauge theory.

\item{\textbf{Case $\mathbf{k=4}$, $\mathbf{G_{8}}$:} }
The CB  and charge lattice can be computed with \eqref{eq:CB_E6} and \eqref{eq:states_E6} respectively with:
\begin{equation}
w=\left(c_{E_6\subset E_7}\right)^{3}
\end{equation}
The theory is a rank 2 SCFT with CB  $\cc^3/G_{8}$ where $G_{8}$ is the ECCRG with degrees 8 and 12. This theory is believed to be the same theory as the exceptional S-fold SCFT of type $E_6$ with $k=4$ studied in Section \ref{sec:G8}. The rank-1 theory supported on the codimension-1 singularity has a $\zz_2$ 1-form symmetry. Then following the same arguments as in Section \ref{sec:G8} we find that this theory must be a discrete gauging of free $U(1)^2$ $\NN=4$ gauge theory.
\item{\textbf{Case $\mathbf{k=6}$, $\mathbf{G_{26}}$:} }
The CB  and charge lattice can be computed with \eqref{eq:CB_E6} and \eqref{eq:states_E6} respectively with:
\begin{equation}
w=\left(c_{E_7}\right)^{3}
\end{equation}
The theory is a rank 3 SCFT with CB  $\cc^3/G_{26}$ where $G_{26}$ is the ECCRG with degrees 6,12 and 18. Performing the same computations as in the $k=3$ case we find that this theory must be a discrete gauging of free $U(1)^3$ $\NN=4$ gauge theory as well.
\end{itemize}

We argued that all the exceptional S-fold theories of type $E_7$ are not interacting SCFTs but rather discrete gauging of free theories. In Section \ref{sec:bound} we will give additional evidence for this claim by showing that it is not possible to define a consistent charge lattice on the CB of these theories.

\subsection{S-folds from the $(2,0)$ $E_8$ theory}
\label{sec:E8}

\begin{table}
\centering
\begin{tabular}{|c|c|c|c|c|}
\hline
\multicolumn{5}{|c|}{$E_8$}\\
\hline
$k$ & $w$ & $r$ & ECCRG & $\Delta_i $\\
\hline
3 & $\left(c_{E_8} \right)^{10}$ & 4 & $G_{32}$ & $\{12,18,24,30\}$ \\
\hline
4 & $\left( c_{E_8} (s_7 s_8)^{-1}  c_{E_8} c_{E_8} (s_7 s_8)^{-1}  \right)^6$ & 4 & $G_{31}$ & $\{8,12,18,24\} $\\
\hline
6 & $\left(c_{E_8} \right)^5$ & 4 & $G_{32}$ & $\{12,18,24,30\} $\\
\hline
\end{tabular}
\caption{Elements $w\in \mathcal{W} \left[E_8\right]$ that characterize the S-fold actions for the S-fold theories of type $E_8$ through \eqref{eq:CB_E6} and \eqref{eq:states_E6}. Here $c_{E_8}$ is the Coxeter element of $E_8$, $s_i$ are the reflection along the $i$-th simple root, $r$ is the rank of the S-fold theory and in the fourth and fifth columns the exceptional complex crystallographic reflection group (ECCRG) associated to the S-fold theories and its degrees $\Delta_i$ are reproduced.}
\label{fig:wE8}
\end{table}

In this section we consider the exceptional S-fold theories obtained from the $(2,0)$ six dimensional theory of type $E_8$. 
Our main result is that the exceptional S-folds SCFTs obtained from the $E_8$ theories with $k=3,6$ are discrete gauging of free $U(1)^r$ $\NN=4$ gauge theory, where $r$ is the rank of the theory, see Table \ref{fig:wE8}.
On the other hand, the exceptional S-fold SCFT of type $E_8$ with $k=4$, also known as the $G_{31}$ theory, passes all consistency checks, therefore we expect it to be a non-trivial interacting $\NN=3$ SCFT. Considering also our results for the exceptional S-fold theories of type $E_6$ and $E_7$ the $G_{31}$ theory is the only exceptional S-fold SCFT of type $E$ which is a proper interacting theory. We also compute the 1-form symmetry group of the $G_{31}$ theory and find it to be trivial.

We work in a basis of the algebra $E_8$ given by simple roots $\alpha_i$ such that the Cartan matrix is the one in Figure \ref{fig:cartanE}. The reflections along the simple roots are denoted as $s_i$ and the corresponding Coxeter element of $E_8$ is:
\begin{equation}
\settowidth\mylen{$-1$}
c_{E_8} =   \prod_{i=1}^{8} s_i  =
\left(\begin{array}{*{8}{wc{\mylen}}}
 0 & 0 & 0 & 0 & 1 & 0 & -1 & -1 \\
 1 & 0 & 0 & 0 & 1 & 0 & -1 & -1 \\
 0 & 1 & 0 & 0 & 1 & 0 & -1 & -1 \\
 0 & 0 & 1 & 0 & 1 & 0 & -1 & -1 \\
 0 & 0 & 0 & 1 & 1 & 0 & -1 & -1 \\
 0 & 0 & 0 & 0 & 1 & 0 & -1 & 0 \\
 0 & 0 & 0 & 0 & 0 & 1 & -1 & 0 \\
 0 & 0 & 0 & 0 & 1 & 0 & 0 & -1 \\
\end{array}
\right)
\end{equation}
which satisfies:
\begin{equation}
\left(c_{E_8}\right)^{30} = \mathbb{1}
\end{equation}
We also report the explicit expressions for $s_7$ and $s_8$, which are used in defining the elements $w$, see Table  \ref{fig:wE8}.
\begin{equation}
\settowidth\mylen{$-1$}
s_7 = 
\left(\begin{array}{*{8}{wc{\mylen}}}
 1 & 0 & 0 & 0 & 0 & 0 & 0 & 0 \\
 0 & 1 & 0 & 0 & 0 & 0 & 0 & 0 \\
 0 & 0 & 1 & 0 & 0 & 0 & 0 & 0 \\
 0 & 0 & 0 & 1 & 0 & 0 & 0 & 0 \\
 0 & 0 & 0 & 0 & 1 & 0 & 0 & 0 \\
 0 & 0 & 0 & 0 & 0 & 1 & 0 & 0 \\
 0 & 0 & 0 & 0 & 0 & 1 & -1 & 0 \\
 0 & 0 & 0 & 0 & 0 & 0 & 0 & 1 \\
\end{array}
\right),
\qquad
s_8 =
\left(\begin{array}{*{8}{wc{\mylen}}}
 1 & 0 & 0 & 0 & 0 & 0 & 0 & 0 \\
 0 & 1 & 0 & 0 & 0 & 0 & 0 & 0 \\
 0 & 0 & 1 & 0 & 0 & 0 & 0 & 0 \\
 0 & 0 & 0 & 1 & 0 & 0 & 0 & 0 \\
 0 & 0 & 0 & 0 & 1 & 0 & 0 & 0 \\
 0 & 0 & 0 & 0 & 0 & 1 & 0 & 0 \\
 0 & 0 & 0 & 0 & 0 & 0 & 1 & 0 \\
 0 & 0 & 0 & 0 & 1 & 0 & 0 & -1 \\
\end{array}
\right)
\end{equation}

\begin{itemize}
\item{\textbf{Case $\mathbf{k=3 \text{ or } 6}$, $\mathbf{G_{32}}$:} }
The exceptional S-folds of type $E_8$ with $k=3$ and $k=6$ give rise to the same field theory. 
The CB  and charge lattice can be computed with \eqref{eq:CB_E6} and \eqref{eq:states_E6} respectively with:
\begin{equation}
w= \left\{
\begin{array}{ll}
\left(c_{E_8}\right)^{10}, \qquad\qquad &k=3
\\
\left(c_{E_8}\right)^{5}, \qquad\qquad &k=6
\end{array}
\right.
\end{equation}
The CB  is $\cc^4/G_{32}$ with $G_{32}$ the ECCRG with degrees 12,18,24 and 30. There is only one codimension-1 singularity up to $G_{32}$ transformation. The transverse slice to this singularity is $\cc/\zz_3$ and the 1-form symmetry group of the theory supported on this singularity is $\zz_3$. There is no rank-1 $\NN=2$ theory compatible with a $\cc/\zz_3$ CB  and with a $\zz_3$ 1-form symmetry group, therefore this singularity must be empty. Then the $G_{32}$ theory itself must be a discrete gauging of free $U(1)^4$ $\NN=4$ gauge theory.

\item{\textbf{Case $\mathbf{k=4}$, $\mathbf{G_{31}}$:} }
The CB  of the S-fold theory of type $E_8$ with $k=4$ can be computed with \eqref{eq:CB_E6}  where:
\begin{equation}
w = \left(c_{E_8}\left(s_7 s_8\right)^{-1} c_{E_8} c_{E_8}\left(s_7 s_8\right)^{-1}\right)^6
\end{equation}
The CB  is $\cc^4/G_{31}$ where $G_{31}$ is the ECCRG with degrees 8,12,20 and 24. Notice that in order for the S-fold theory to be rank 4, $w$ must have a four eigenvalues $i$.
$w$ is real, therefore it must also have four eigenvalues $-i$, and thus $w^2 = - \mathbb{1}$. Then one can consider the states:
\begin{equation}	\label{eq:statesG31}
\overline{|\alpha,(p, q)\rangle}_{\text{short}}=\frac{1}{\sqrt{2}} \sum_{t=0}^1\left|w^t \cdot \alpha ;\left(\rho_4\right)^t \cdot(p, q)\right\rangle
\end{equation}
where $\alpha$ is a root of $E_8$. The states \eqref{eq:statesG31} are invariant under the S-fold action for any $\alpha$ and $(p,q)$, therefore we consider the charge lattice $\Gamma$ spanned by the charges of the states \eqref{eq:statesG31}. A basis for this lattice is given by the charges $q_i$ of the states $\overline{|\alpha_i,(1, 0)\rangle}_{\text{short}}$ obtained from the W-bosons associated with the simple roots $\alpha_i$ of $E_8$ with $i=1,\dots,8$:
\begin{equation}
\settowidth\mylen{$-1$}
\begin{array}{c*{16}{wc{\mylen}}}
 q_1 = \displaystyle \frac{1}{\sqrt{2}}&(  1, & 0, & 0, & 0, & 0 ,& 0, & 0, & 0, & 0 ,& 1 ,& 0, & 1 ,& 0, & 0, & 0 ,& 0 ) \\
 q_2 =  \displaystyle \frac{1}{\sqrt{2}}&( 0, & -1, & 1, & -2, & 0, & -3 ,& 0, & -4, & 0, & -5 ,& 0 ,& -4 ,& 0 &, -2 & 0, & -2) \\
q_3 = \displaystyle \frac{1}{\sqrt{2}} &(  0, & 1, & 0, & 2, & 1, & 3, & 0, & 4 ,& 0 ,& 4 ,& 0, & 3, & 0, & 1, & 0, & 2) \\
q_4 = \displaystyle \frac{1}{\sqrt{2}} &(  0, & 0, & 0, & -1, & 0 ,& -2, & 1, & -2, & 0, & -2, & 0, & -1, & 0, & 0, & 0, & -1) \\
q_5 = \displaystyle \frac{1}{\sqrt{2}}&(  0, & -1, & 0, & -1, & 0, & -1, & 0, & -2, & 1, & -3, & 0, & -2, & 0, & -1, & 0, & -2) \\
q_6 = \displaystyle \frac{1}{\sqrt{2}} &(  0, & 0, & 0, & 1, & 0, & 1, & 0, & 2, & 0, & 3, & 1, & 2, & 0, & 1, & 0, & 2) \\
q_7 =  \displaystyle \frac{1}{\sqrt{2}}&(  0, & 1, & 0, & 1, & 0, & 1, & 0 ,& 0, & 0, & 0, & 0 ,& 0, & 1, & 0, & 0, & 0) \\
q_8 =  \displaystyle \frac{1}{\sqrt{2}}&(  0, & 1, & 0, & 1, & 0, & 2, & 0, & 3, & 0, & 4, & 0, & 2, & 0, & 1, & 1, & 2) \\
\end{array}
\end{equation}

The Dirac pairing $J_{G_{31}}$ in this basis  can be computed with \eqref{eq:J_E6}:
\begin{equation}
\settowidth\mylen{$-1$}
J_{G_{31}} = 
\left(\begin{array}{*{8}{wc{\mylen}}}
 0 & 0 & 0 & 1 & -1 & -1 & 1 & 1 \\
 0 & 0 & 0 & 0 & 0 & 1 & 0 & -1 \\
 0 & 0 & 0 & -1 & 1 & -1 & 1 & 0 \\
 -1 & 0 & 1 & 0 & 0 & 0 & -1 & 0 \\
 1 & 0 & -1 & 0 & 0 & 0 & 0 & 1 \\
 1 & -1 & 1 & 0 & 0 & 0 & 0 & -1 \\
 -1 & 0 & -1 & 1 & 0 & 0 & 0 & 0 \\
 -1 & 1 & 0 & 0 & -1 & 1 & 0 & 0 \\
\end{array}
\right)
\end{equation}
The order of the 1-form symmetry group of the $G_{31}$ is given by the absolute value of the Pfaffian of $J_{G_{31}}$:
\begin{equation}
\left| G^{(1)} \right| = \left| \pf[J_{G_{31}}] \right | = 1
\end{equation}
Therefore the $G_{31}$ theory has trivial 1-form symmetry. 

Let us now consider the stratification of the CB  $\cc^4/G_{31}$. There is one codimension-1 singularity up to $G_{31}$ transformations and the sublattice of $\Gamma$ corresponding to the states becoming massless on this singularity is compatible with $SU(2)$ $\NN=4$ SYM. The transverse slice to this singularity is $\cc/\zz_2$ which is compatible with the CB  of $SU(2)$ $\NN=4$. This stratification is consistent with the central charge formulae of \cite{Martone:2020nsy}, as already checked in \cite{Kaidi:2022lyo}. We briefly review the relevant computations here for ease of readibility.
The Shapere-Tachikawa formula with $a=c$ allow us to compute the central charges:
\begin{equation}
2c = \sum_{j=1}^r \Delta_j-\frac{r}{2} = 8 + 12+20+24 - 2 = 62
\end{equation}
While the formulae of \cite{Martone:2020nsy} relate the central charges with data of the theory supported on the codimension-1 singularity. For the $G_{31}$ theory this formula reads:
\begin{equation}
12 c=3 r+ \Delta^{\mathrm{sing}} b = 12 + 120 b
\end{equation}
where $\Delta^{\mathrm{sing}} = \text{lcm} (\Delta_i)$ is the scaling dimension of the codimension-1 singularity and $b$ is associated to the data of the rank-1 theory supported on this singularity by \eqref{eq:b_martone}. Solving for $b$ we find:
\begin{equation}
b=3
\end{equation}
which is consistent with having $SU(2)$ $\NN=4$ SYM as the theory supported on the codimension-1 singularities.
\end{itemize}

To summarize, we have found that the exceptional S-fold theories of type $E_8$ with $k=3,6$ are discrete gauging of free theories. The exceptional S-fold theory of type $E_8$ with $k=4$, denoted as the $G_{31}$ theory, passes all the consistency checks that we have at our disposal. We are then lead to claim that the $G_{31}$ theory is the only exceptional S-fold theory of type $E_{6,7,8}$ that is a proper interacting SCFT. We have also found that the 1-form symmetry group of the $G_{31}$ theory is trivial.

\section{Charge lattices in $N=2$ SCFTs with $\varkappa \neq \{1,2\}$}
\label{sec:bound}
One of the main features of S-fold SCFTs it that they are maximally strongly coupled theories. This means that whenever a charged state become massless then another charged state which is mutually non local with respect to the first one becomes massless as well. Then at any non-generic point of the CB  the low energy theory is strongly coupled and does not admit a conventional lagrangian description. 
On the one hand this fact renders the study of S-fold SCFTs challenging, because the only vacua where perturbative techniques are viable are the most generic points of the CB , where the low energy theory is simply $U(1)^r$ with no massless charged states. 
On the other hand, as was shown in \cite{Cecotti:2021ouq}, maximally strongly coupled theories have to satisfy a series of non-trivial constraints, and are quite restricted as a result. Motivated by the results of \cite{Cecotti:2021ouq} in this Section we study the charge lattices of 
a large class of $\NN=2$ SCFTs that are maximally strongly coupled,
namely $\NN=2$ SCFTs with characteristic dimension $\varkappa \neq \{1,2\}$. All the regular and exceptional S-fold SCFTs belong to this class of theories except for the known cases where SUSY enhances to $\NN=4$ \cite{Aharony:2016kai}.
This Section is meant to be readable independently from the rest of this paper, therefore there are some repetitions between this and the other Sections.

For the rest of this section we only consider interacting rank-$r$ SCFTs with $\varkappa \neq \{1,2\}$. Our main results are:

\setcounter{claim}{0}
\begin{claim}
The order of the 1-form symmetry group $G^{(1)}$ of an $\NN=2$ rank-2 SCFT with $\varkappa \neq \{1,2\}$ satisfies $1\leq \left|G^{(1)}\right| \leq 4$. The upper bound can only be saturated by stacks of lower rank theories.
\end{claim}

\begin{claim}
An $\NN=2$ SCFT with $\varkappa\neq\{1,2\}$ and rank $r\geq2$  that is not a stack of lower rank theories must have at least one codimension-1 singularity that supports (a discrete gauging of) $\NN=2^*$ $SU(2)$ SYM.
\end{claim}

We also apply the techniques that we develop to the exceptional S-fold theories, which provides an independent argument for claiming that most of these theories are not interacting SCFTs.

Let us review the definitions and results of \cite{Cecotti:2021ouq} that will be useful in this Section.
The characteristic dimension of an $\NN=2$ SCFT is defined as follows. Write the degrees of CB  invariants $\Delta_i$ as:
\begin{equation}
	( \Delta_1, \dots, \Delta_N ) = \lambda (d_1, \dots, d_2), \qquad d_i \in \zz, \quad \text{gdc}(d_1, \dots, d_N) = 1
\end{equation}  
Then the characteristic dimension is defined as:
\begin{equation}
	\varkappa=\frac{1}{\left\{\lambda^{-1}\right\}}
\end{equation}
where $\{x\}$ is defined as the unique real number such that $\{x\} = x \text{ mod }1$ and $0< \{x\} \leq 1$. 
The characteristic dimension can only take eight values:
\begin{equation}
\varkappa \in \{ 1, 6/5, 4/3, 3/2, 2,3,4,6 \}
\end{equation}
 An SCFT with $\varkappa \neq \{1,2\}$ is maximally strongly coupled and for any state with charge $q$, central charge $Z[q] \neq 0$ there is another state with charge $p$ and central charge $Z[p]=\zeta Z[q]$ where:
\begin{equation}	\label{eq:Dirac_H}
\settowidth\mylen{$111111111$}
\zeta = \left\{
\begin{array}{*{1}{wl{\mylen}}l}
e^{2\pi i/3} \qquad & \varkappa = 3, 3/2 \\
i \qquad  &\varkappa = 4, 4/3 \\
e^{2\pi i/6}  &\varkappa = 6, 6/5 \\
\end{array}
\right.
\end{equation}
therefore the charge lattice $\Gamma$ is mapped to a lattice $\zz[\zeta]^r$ by the central charge $Z$. The Dirac pairing between two charges $p$ and $q$ can be written as:
\begin{equation}
\langle q,p \rangle = \frac{1}{\zeta - \overline{\zeta}} \left( H(q,p) - H(p,q)
\right)
\end{equation}
here and in the remainder of this Section, with an abuse of notation, we denote with the same symbol the electromagnetic charges in $\Gamma$ and the corresponding element in the lattice  $\zz[\zeta]^r$.  Here $H$ is a positive definite Hermitian form:
\begin{equation}
H: \overline{\zz[\zeta]^r} \times \zz[\zeta]^r \to \zz[\zeta]
\end{equation}
An important implication of having $\varkappa \neq \{1,2\}$ is that when a state with charge $q$ becomes massless then also a state with charge $p=\zeta q$ becomes massless. Then we can always choose a basis of the lattice of charges becoming massless at a codimension-1 singularity that is of the form $\{ q, \zeta q\}$.

\subsection{1-form symmetries of rank-2 $\NN=2$ SCFTs with $\varkappa \neq \{1,2\}$}
Consider a rank-2 $\NN=2$ SCFTs with $\varkappa \neq \{1,2\}$. We choose a basis of the charge lattice $\Gamma$ of the form $\{ q_1, \zeta q_1, q_2, \zeta q_2\}$:
\begin{equation}
\Gamma = \text{Span}_{\zz} \{ q_1, \zeta q_1, q_2, \zeta q_2\}
\end{equation}
Were $q_1$ and $\zeta q_1$ become massless at some codimension-1 singularity $\mathcal{H}_1$ and $q_2$ and $\zeta q_2$ become massless at some other codimension-1 singularity $\mathcal{H}_2$. $q_1$ and $q_2$ must be linearly independent for $\Gamma$ to have dimension 4. 

An interesting quantity to consider is the absolute value of the Pfaffian of the Dirac pairing $J$, which is an invariant of the charge lattice and intuitively tells us how sparse the charge lattice is. More precisely in \cite{Argyres:2022kon, DelZotto:2022ras} it was shown that this quantity is equal to the order of the 1-form symmetry group, which in turn is related to how much the charge lattice can be refined without breaking the Dirac quantization condition. The number of charges that can be added in the fundamental domain of the charge lattice while preserving the Dirac quantization condition is given by $\left| \pf[J] \right|$ minus 1.

Consider the rank-1 theory $\mathcal{T}_i$ supported on the codimension-1 singularity $\mathcal{H}_i$. Its charge lattice is spanned by $\{ q_i, \zeta q_i\}$ and the Dirac pairing $J_{\mathcal{H}_i}$ is such that:
\begin{equation}
\left| \pf[J_{\mathcal{H}_i}] \right| =
\left| \text{Pf} \left(
\begin{array}{cc}
0 & \langle q_i, \zeta q_i \rangle \\
- \langle q_i, \zeta q_i \rangle & 0\\
\end{array}
\right) \right|
= \left| \langle q_i, \zeta q_i \rangle \right| = H(q_i, q_i)
\end{equation}
where in the last equality we used \eqref{eq:Dirac_H}. $\mathcal{T}_i$ is an $\NN=2$ rank-1 SCFT, therefore the 1-form symmetry is either $\zz_2$, if this theory is (a discrete gauging of) $\NN=2^*$ $SU(2)$ SYM, or trivial in any other case. Therefore we found:
\begin{equation}	\label{eq:Hi_rank2}
H(q_i, q_i) = \left\{
\begin{split}
&2 \qquad \mathcal{T}_i \text{ is (a discrete gauging of) $\NN=2^*$ $SU(2)$ SYM}\\
&1 \qquad \text{otherwise}
\end{split}
\right.
\end{equation}

Now let us compute the Pfaffian of the Dirac pairing $J^{(2)}$ of the rank-2 theory itself. We find:
\begin{equation}
\begin{split}
\left|\pf[J^{(2)}] \right|=& \left| \text{Pf} \left(
\begin{array}{cccc}
0 & \DD( q_1, \zeta q_1 ) & \DD( q_1, q_2 ) & \DD( q_1,\zeta q_2 )
\\
\DD(\zeta q_1,q_1) & 0 & \DD(\zeta q_1,q_2) & \DD(\zeta q_1,\zeta q_2)
\\
\cdots & \cdots & 0 & \DD(q_2,\zeta q_2)
\\
\cdots & \cdots & \DD(\zeta q_2,q_2) & 0
\end{array}
\right)
\right|
\\
=&\left| \DD( q_1, \zeta q_1 ) \DD(q_2,\zeta q_2) - \DD( q_1, q_2 ) \DD(\zeta q_1,\zeta q_2)+\DD(\zeta q_1,q_2) \DD( q_1,\zeta q_2 ) \right|
\\
=& H(q_1,q_1) H(q_2,q_2)- \left|H(q_1,q_2)\right|^2
\end{split}
\end{equation}
where we dropped the absolute value in the last line because the Cauchy-Schwarz inequality ensures that the last expression is positive. We are now able to determine upper and lower bounds for this quantity:
\begin{equation}	\label{eq:r2_bound_raw}
1\leq \left|\pf[J^{(2)}] \right| \leq  H(q_1,q_1) H(q_2,q_2) \leq 4
\end{equation}
The first inequality holds because the Dirac pairing is integer and non-degenerate, while the last inequality follows from the analysis of the rank-1 theories supported on the codimension-1 singularities \eqref{eq:Hi_rank2}. The inequality:
\begin{equation}	\label{eq:CS_applied}
\left|\pf[J^{(2)}] \right| \leq  H(q_1,q_1) H(q_2,q_2)
\end{equation}
is saturated only if $H(q_1,q_2)$ vanishes. Then in order to have a rank-2 SCFT with $\varkappa \neq\{1,2\}$ with $\left|\pf[J^{(2)}] \right|=4$ it is necessary that for every choice of codimension-1 singularities $\mathcal{H}_1, \mathcal{H}_2$ the theories supported on the singularities is (a discrete gauging of ) $\NN=2^*$ SYM and  that $H(q_1,q_2)=0$. This means that, for every choice of $\mathcal{H}_1, \mathcal{H}_2$, the charges becoming massless on $\mathcal{H}_1$ are mutually local with respect to the charges becoming massless on $\mathcal{H}_2$. The rank-2 theory then must be a stack of the rank-1 theories supported on $\mathcal{H}_i$. 
We are not interested in theories that are stacks of lower rank theories, therefore we can drop the equal sign in \eqref{eq:CS_applied}, and \eqref{eq:r2_bound_raw} reduces to:
\begin{equation}	\label{eq:r2_bound}
1\leq \left|\pf[J^{(2)}] \right| \leq  3
\end{equation}
As already discussed the absolute value of the Pfaffian of the Dirac pairing is equal to the order of the 1-form symmetry group, therefore we find our first claim:
\setcounter{claim}{0}
\begin{claim}
The order of the 1-form symmetry group $G^{(1)}$ of an $\NN=2$ rank-2 SCFT with $\varkappa \neq \{1,2\}$ satisfies $1\leq \left|G^{(1)}\right| \leq 4$. The upper bound can only be saturated by stacks of lower rank theories.
\end{claim}
Consider now a theory where $H(q_i,q_i)=1$ for every choice of $\mathcal{H}_i$. 
As we just discussed equation \eqref{eq:CS_applied} is only saturated for rank-2 theories that are stacks of rank-1 theories. Then for a $\NN=2$ rank-2 SCFT with $\varkappa \neq \{1,2\}$ that is not a stack of lower rank theories we have:
\begin{equation}	\label{eq:r2_inconsistency}
1\leq \left|\pf[J^{(2)}] \right| < 1
\end{equation}
This is a contradiction and signals that on such a CB  it is not possible to define a consistent charge lattice. We can then formulate our second claim in the case of rank-2:
\renewcommand\theclaim{\Alph{claim}'}
\begin{claim}
A rank-2 $\NN=2$ SCFT with $\varkappa\neq\{1,2\}$ that is not a stack of lower rank theories must have at least one codimension-1 singularity that supports (a discrete gauging of) $\NN=2^*$ $SU(2)$ SYM.
\end{claim}
\renewcommand\theclaim{\Alph{claim}}
As we will see in the next Section this second claim generalizes to arbitrary rank. 

In the context of $\NN=3$ exceptional S-fold SCFT the second claim already rules out some of the CB  geometries. The most straightforward to study are the $G_5$ theory that we studied in Section \ref{sec:G5} as well as the $G_4$ theory that can be constructed from the $D_4$ $(2,0)$ six-dimensional theory with an S-fold procedure in the presence of an outer automorphism twist. Both these theories are maximally strongly coupled and only have codimension-1 singularities with a transverse slice $\cc/\zz_3$, which can not support a discrete gauging of $\NN=2^*$ $SU(2)$ SYM.

The $G_8$ theory, studied in Section \ref{sec:G8}, is more subtle. There is one codimension-1 singularity with transverse slice $\cc/\zz_4$. Our second claim then imposes that the theory supported on this singularity is a $\zz_4$ gauging of $\NN=4$ $SU(2)$ SYM. On the other hand the analysis of the central charges with the formulae of \cite{Martone:2020nsy} is not consistent with this choice, as already computed in \cite{Kaidi:2022lyo} and as we discussed in Section \ref{sec:G8}.
To summarize, we find that the exceptional S-fold theories $G_{4}$, $G_{5}$ and $G_{8}$  can not be interacting SCFTs because it is not possible to define a consistent charge lattice on their CB. Therefore these theories must be discrete gaugings of free $U(1)^r$ $\NN=4$ gauge theories, which is the only other possibility.

\subsection{A constraint for the stratification of $\NN=2$ SCFTs}
Consider a rank-$r$ $\NN=2$ SCFT with $\varkappa\neq\{1,2\}$ such that the rank-1 theories supported on all codimension-1 singularities are SCFTs with trivial 1-form symmetries. We can choose a basis of the charge lattice $\Gamma$ such that:
\begin{equation}
\Gamma = \text{Span}_{\zz} \{q_1, \zeta q_1, \dots, q_r, \zeta q_r \}
\end{equation}
where $q_i$ and $\zeta q_i$ become massless at some codimension-1 singularity $\mathcal{H}_i$ and generate the charge lattice of the rank-1 theory supported there. Then we have:
\begin{equation}
H(q_i, q_i) = 1 \qquad \forall i=1,\dots,r
\end{equation}
because $H(q_i,q_i)$ is equal to the order of the 1-form symmetry group of the theory supported on $\mathcal{H}_i$, which is trivial by hypotesis. The Chaucy-Schwarz inequality together with the fact that all the $q_i$ are linearly independent imposes:
\begin{equation}
\left| H(q_i, q_j) \right|^2 < H(q_i,q_i) H(q_j,q_j) = 1 \qquad \forall i\neq j
\end{equation}
On the other hand $\left| H(q_i, q_j) \right|^2$ must be an integer because it can be written as an integer linear combination of Dirac pairings:
\begin{equation}
\left| H(q_i, q_j) \right|^2 = \left\langle q_i, q_j\right\rangle\left\langle\zeta q_i, \zeta q_j\right\rangle
-
\left\langle\zeta q_i, q_j\right\rangle\left\langle q_i, \zeta q_j\right\rangle
\end{equation}
Then $\left| H(q_i, q_j) \right|^2$ must vanish and $H(q_i, q_j)$ vanishes as well as a consequence. The resulting Dirac pairing matrix is block diagonal with only $2\times 2$ blocks:
\begin{equation}
J^{(r)} = \text{diag} \left\{
\left(\begin{array}{cc} 0 & 1 \\ -1 & 0 \end{array}\right),
\left(\begin{array}{cc} 0 & 1 \\ -1 & 0 \end{array}\right),
\dots,
\left(\begin{array}{cc} 0 & 1 \\ -1 & 0 \end{array}\right)
\right\}
\end{equation}
This is the case for every choice of codimension-1 singularities $\mathcal{H}_i$, therefore the states becoming massless at any singularity $\mathcal{H}_i$ are mutually local with respect to the states becoming massless at any other singularity. Then the rank-$r$ theory must be a stack of $r$ rank-1 theories. Therefore in order to have a rank-$r$ SCFT with $\varkappa \neq \{1,2\}$ that is not a stack of lower rank theories, one must allow for codimension-1 singularities $\mathcal{H}_i$ that support (a discrete gauging of) $\NN=2^*$ $SU(2)$ SYM, which would imply that $H(q_i,q_i)=2$. This is our second claim:
\setcounter{claim}{1}
\begin{claim}
An $\NN=2$ SCFT with $\varkappa\neq\{1,2\}$ and rank $r\geq2$ that is not a stack of lower rank theories must have at least one codimension-1 singularity that supports (a discrete gauging of) $\NN=2^*$ $SU(2)$ SYM.
\end{claim}

This claim rules out the exceptional S-fold theories $G_{25}$ and $G_{32}$ as possible interacting SCFTs alongside the rank-2 cases discussed in the previous section. Indeed in both cases all codimension-1 singularity have transverse slice $\cc/\zz_3$ which can not support a discrete gauging of $\NN=2^*$ $SU(2)$ SYM, and both theories are maximally strongly coupled and can not have IR-free theories supported on any singularity. In the case of the $G_{26}$ theory, studied in Section \ref{sec:E7}, there is only one singularity with transverse slice $\cc/\zz_2$ which could support $\NN=2^*$ $SU(2)$ SYM, while the other singularity has transverse slice $\cc/\zz_3$. However, as computed in \cite{Kaidi:2022lyo} and discussed above, this choice is inconsistent with central charge formulae of \cite{Martone:2020nsy}. To summarize, we find that the $G_{25}$, $G_{26}$ and $G_{32}$ theories can not be interacting SCFTs because it is not possible to define a consistent charge lattice on their CB. Therefore these theories must be discrete gaugings of free theories, which is the only other possibility.

The only exceptional S-fold SCFT obtained from the $(2,0)$ $E_n$ six-dimensional theories that satisfies our constraint is the $G_{31}$ theory. This is the SCFT obtained from the $E_8$ $(2,0)$ theory with an exceptional S-fold of order $k=4$. The CB  is $\cc^4/G_{31}$ and has one codimension-1 singularity with transverse slice $\cc/\zz_2$. Our constraint then imposes that the rank-1 theory supported on this singularity is $SU(2)$ $\NN=4$ SYM, which is consistent with the analysis of the central charges performed in \cite{Kaidi:2022lyo}. We claim that this theory is a proper interacting rank-4 SCFT, but it must be noted that the possibility of having a discrete gauging of $U(1)^4$ $\NN=4$ gauge theory is also consistent with all the constraints that are available. To solve this ambiguity it would be desirable to analyze the spectrum of charged operators directly from the M-theory setup of \cite{Garcia-Etxebarria:2016erx}, but we leave this problem to future work.

\subsection{Other exceptional S-folds}
In this paper we have analyzed the exceptional S-fold theories obtained from the $(2,0)$ $E_n$ six-dimensional theories, but it is also possible to define similar M-theory setups involving the $(2,0)$ $D_n$ theories, with or without outer automorphism twists, and generalizations to non-simply laced algebras have also been considered. We leave the detailed analysis of the charge lattices of the resulting theories to future work, but we can easily check if the CBs of such theories, computed in  \cite{Kaidi:2022lyo}, satisfy our consistency conditions.

The CB of the S-fold SCFTs obtained from the $(2,0)$ $D_n$ theories are $\cc^r/G(k,m,r)$ for $k=4,6$ and $m$ a divisor of $k$. In all these cases there is at least one codimension-1 singularity with transverse slice $\cc/\zz_2$ which can support $SU(2)$ $\NN=4$ SYM, therefore these theories satisfy our consistency condition. 

The case of non-simply laced algebra generates one CB  geometry that does not appear in the exceptional S-fold theories of type $E_n$ and $D_n$, namely $\cc^2/G_{12}$. Here $G_{12}$ is the exceptional complex reflection group with degrees 6 and 8. In this geometry there are codimension-1 singularity with transverse slice $\cc/\zz_2$ that can support $SU(2)$ $\NN=4$ SYM, therefore this theory satisfies our consistency checks.

As already discussed in \cite{Kaidi:2022lyo} there are four possible geometries associated to ECCRGs that do not appear in any known construction but are consistent CBs for putative $\NN=3$ SCFTs. These geometries are $\cc^3/G_{24}$, $\cc^4/G_{29}$, $\cc^5/G_{33}$ and $\cc^6/G_{34}$ where $G_i$ are ECCRGs. In all cases there are codimension-1 singularities that have transverse slice $\cc/\zz_2$ and could support $SU(2)$ $\NN=4$ SYM, therefore all these CBs satisfy our consistency condition.

\section{Conclusion}
\label{sec:concl}
In this paper we studied the exceptional S-fold SCFTs discovered in \cite{Garcia-Etxebarria:2016erx} and their associated charge lattices. In Section \ref{sec:exceptional} we analyzed explicitly all the exceptional $\sfolds(E_{6,7,8},k)$, computing their charge lattices by generalizing the techniques of \cite{Kaidi:2022lyo}. 
Furthermore we considered the sublattices of charges that become massless at codimension-1 singularities of the CB . 
These sublattices must correspond to the charge lattices of some rank-1 $\NN=2$ SCFT because S-fold SCFTs are maximally strongly coupled. Moreover this match must be consistent with the constraints on the central charges from  \cite{Martone:2020nsy}.
We find that the charge lattices of most of the $\sfolds(E_{6,7,8},k)$ do not satisfy these constraints, therefore they can not be interacting SCFTs.
For exceptional $\sfolds(E_{6,7,8},k)$ the only theory that admits a consistent charge lattice is the S-fold of type $E_8$ with $k=4$, called the $G_{31}$ theory. 
Thus we claimed that the $G_{31}$ theory is an interacting SCFT, while all the other $\sfolds(E_{6,7,8},k)$ are discrete gauging of free theories.

In this Section \ref{sec:bound} we provided additional evidence for this claim by studying the charge lattice of $\NN=2$ SCFTs with characteristic dimension $\varkappa \neq \{1,2\}$. By exploiting the results and the formalism developed in \cite{Cecotti:2021ouq} we computed an upper bound for the 1-form symmetries of rank-2 theories with $\NN=2$, denoted as \textbf{Claim} \ref{ClaimA} throughout this paper, and we found a consistency constraints for the CB  stratification for such SCFTs at any rank, denoted as \textbf{Claim} \ref{ClaimB} throughout this paper.
When applied to the case of exceptional S-fold SCFTs this constraint in combination with other constraints from \cite{Martone:2020nsy} corroborates our results spelled out above.

There are multiple directions our work can be extended towards. 
As we already commented in main body of this paper the $G_{31}$ theory passes all our consistency checks, but this does not guarantee that this theory is indeed an interacting SCFT and it may be a discrete gauging of a free theory.
A possible way to determine whether this is the case would be to compute the 2-form symmetries of this theory directly from the M-theory construction. Indeed if the 2-form symmetry group is not trivial and can be gauged then the $G_{31}$ theory should be a discrete gauging of some “parent” theory. This is the approach  adopted in \cite{Aharony:2016kai} to understand the presence of discrete gauging in regular S-folds. 

Our procedure could be applied to study the generalized symmetries in various classes of 4d SCFTs, including the $\sfolds(D_n,k)$, $\NN=2$ S-folds \cite{Apruzzi:2020pmv, Giacomelli:2020jel, Giacomelli:2020gee, Heckman:2020svr} and $\NN=2$ SCFTs with  $\varkappa \neq \{1,2\}$.
In particular in \cite{Argyres:2019ngz} some putative rank-2 $\NN=3$ CB geometries were found to posses complex singularities, which are usually generated by discrete gauging
\footnote{We are grateful to Mario Martone for suggesting us this possibility.}
 \cite{Argyres:2018wxu, Argyres:2016yzz, Bourget:2018ond}. It would be interesting to perform a similar analysis to the one spelled out in this paper in order to understand whether these singularities  can support an interacting theory or not, and therefore arise from discrete gauging.

Our results on $\NN=2$ SCFTs with $\varkappa \neq \{1,2\}$ could be expanded upon in different directions.
It would be nice to generalize \textbf{Claim} \ref{ClaimA} to arbitrary ranks, and understand wether a bound for the 1-form symmetry group exists also at higher ranks. Another interesting perspective would be to relax the condition on the characteristic dimension, and study theories with $\varkappa$ equal to 1 or 2. 
This would require a different set of tools than the ones we used in Section \ref{sec:bound}, for example considering the monodromies around CB  singularities along the lines of \cite{Argyres:2015ffa, Argyres:2015gha,Argyres:2018zay}.
%
%
%

\section*{Acknowledgments}
%
%
We are extremely grateful to Mario Martone and Philippe Argyres for precious comments and for carefully reading the draft.
We are also grateful to Antoine Pasternak for discussions.
The work of A.A. and S.R.  has been supported in part by the Italian Ministero dell'Istruzione, 
Universit\`a e Ricerca (MIUR) and  in part by Istituto Nazionale di Fisica Nucleare (INFN) through the “Gauge Theories, Strings, Supergravity” (GSS) research project.
\pagebreak
\appendix

\section{1-form symmetries of regular S-folds}

In this Appendix we compare the formalism developed in this paper to the 1-form symmetry groups of regular S-fold SCFTs computed in \cite{Etheredge:2023ler,Amariti:2023hev}. In particular we compute the order of the 1-form symmetry group and find agreement with the results in the literature.
The S-fold setup is the one of \cite{Garcia-Etxebarria:2015wns} and reviewed in Section \ref{sec:Sfolds}.

Let us consider an $\sfolds(A_{kN},k)$ with trivial discrete torsion.
A possible choice of basis for the charge lattice is given by fundamental strings ($(1,0)$-strings)  or D1-strings ($(0,1)$-strings), plus their images, stretched between the 1st and $i$-th D3-brane with $i=2,3,\dots,N,N+2$. In the notation introduce above these states are $\overline{|(1,0)\rangle_{1,i} }$ and $\overline{|(0,1)\rangle_{1,i} }$ respectively, with  $i=2,3,\dots,N,N+2$. Their charges are:
\begin{equation}
\begin{split}
Q\left[\overline{|(1,0)\rangle_{1,i} }\right] &= \frac{1}{\sqrt{k}}(1,0; \dots; \overbrace{-1,0}^{i\text{-th}};\dots;  \overbrace{\rho_k\cdot(1,0)}^{N\text{-th}};\dots;  \overbrace{\rho_k\cdot(-1,0)}^{(N+i)\text{-th}}; \dots)
\\
Q\left[\overline{|(0,1)\rangle_{1,i} }\right] &=  \frac{1}{\sqrt{k}}(0,1; \dots; \overbrace{0,-1}^{i\text{-th}};\dots;  \overbrace{\rho_k\cdot(0,1)}^{N\text{-th}};\dots;  \overbrace{\rho_k\cdot(0,-1)}^{(N+i)\text{-th}}; \dots)
\end{split}
\end{equation}
for $i=2,3,\dots,N$ and:
\begin{equation}
\begin{split}
Q\left[\overline{|(1,0)\rangle_{1,N+2} }\right] &= \frac{1}{\sqrt{k}}(1,0; \rho_k^{k-1}\cdot(-1,0);\dots;  \overbrace{\rho_k\cdot(1,0)}^{N\text{-th}};  \overbrace{-1,0}^{(N+2)\text{-th}}; \dots)
\\
Q\left[\overline{|(0,1)\rangle_{1,N+2} }\right] &=  \frac{1}{\sqrt{k}}(0,1; \rho_k^{k-1}\cdot(0,-1);\dots;  \overbrace{\rho_k\cdot(0,1)}^{N\text{-th}};  \overbrace{0,-1}^{(N+2)\text{-th}}; \dots)
\end{split}
\end{equation}
The elements $\rho_k$ of $SL(2,\zz)$ are reported in table \ref{tab:rho}.
In the ordered basis $\mathcal{B}$ for the charge lattice:
\begin{equation}
\mathcal{B} =\left\{\overline{|(1,0)\rangle_{1,N+2} },\overline{|(0,1)\rangle_{1,N+2} },
\left.\overline{|(1,0)\rangle_{1,i} }\right|_{i=2,3,\dots,N},
\left.\overline{|(0,1)\rangle_{1,i} }\right|_{i=2,3,\dots,N} \right\}
\end{equation}
the Dirac pairing is represented by the antisymmetric matrix:
\begin{equation}
J_{k,N}=
\left(
\begin{tabular}{cc|cccc|cccc}
0 & 2 & a & 0 & 0 & \dots & b+1 & 1 & 1 & \dots\\
-2 & 0 & c-1 & -1 & -1 & \dots & d & 0 & 0 & \dots\\
\hline
\multicolumn{2}{c|}{$\begin{gathered} \\ \vdots \\ \; \end{gathered}$} 
&
\multicolumn{4}{c|}{$\begin{gathered} \\ \textbf{0}_{(N-1)\times (N-1)} \\ \; \end{gathered}$} 
&
\multicolumn{4}{c}{$\begin{gathered} \\ M \\ \; \end{gathered}$} 
\\ \hline
\multicolumn{2}{c|}{$\begin{gathered} \\ \vdots \\ \; \end{gathered}$} 
&
\multicolumn{4}{c|}{$\begin{gathered} \\ -M^{\text{T}} \\ \; \end{gathered}$} 
&
\multicolumn{4}{c}{$\begin{gathered} \\ \textbf{0}_{(N-1)\times (N-1)} \\ \; \end{gathered}$} 
\end{tabular}
\right)
\end{equation}
where:
\begin{equation}
M = \left(
\begin{tabular}{ccccc}
2&1&1&\dots&
\\
1&2&1&1&\dots
\\
1&1&\multicolumn{2}{c}{\multirow{2}{*}{$\ddots$}}  &
\\
$\vdots$& & & & 1
\\
&&&1&2
\end{tabular}
\right)
\end{equation}
and:
\begin{equation}
\left(
\begin{tabular}{cc}
a&b \\ c&d
\end{tabular}
\right)
=
\left((\rho_k)^{k-1} \right)^{\text{T}} \cdot 
\left(
\begin{tabular}{cc}
0&1\\-1&0
\end{tabular}
\right)
\end{equation}
As discussed above the absolute value of the Pfaffian of $J_{k,N}$ equals the order of the 1-form symmetry group for the corresponding S-fold SCFTs in the absence of discrete torsion. These 1-form symmetry groups were computed in \cite{Etheredge:2023ler,Amariti:2023hev}:
\begin{equation}
G^{(1)} = \left\{
\begin{split}
\zz_3 \qquad k=3
\\
\zz_2 \qquad k=4
\\
\mathbb{1} \qquad k=6
\end{split}
\right.
\end{equation}
and are independent on the rank $N$. Therefore one expects that:
\begin{equation}	\label{eq:pfaffianSfold}
\left| \Pf \left(J_{k,N}\right) \right| 
= \left\{
\begin{split}
3 \qquad k=3
\\
2 \qquad k=4
\\
1 \qquad k=6
\end{split}
\right.
\qquad \forall \,N
\end{equation}
We have checked numerically that Equation \eqref{eq:pfaffianSfold} holds up to rank $N=100$. It would be interesting to compute $\left| \Pf \left(J_{k,N}\right) \right| $ at arbitrary rank, we leave this problem to future work.

\bibliographystyle{JHEP}
\bibliography{ref}

\end{document}